\documentclass[pra,twocolumn,showpacs,superscriptaddress,floatfix]{revtex4} %4
\newcount\formato \formato=4
\let\a=\alpha \let\b=\beta  \let\g=\gamma  \let\d=\delta \let\e=\varepsilon
\let\z=\zeta  \let\h=\eta   \let\th=\theta \let\k=\kappa \let\l=\lambda
\let\m=\mu    \let\n=\nu    \let\x=\xi     \let\p=\pi    \let\r=\rho
\let\s=\sigma \let\t=\tau   \let\f=\varphi 
\let\ch=\chi  \let\ps=\psi  \let\o=\omega
\let\G=\Gamma \let\D=\Delta \let\Th=\Theta\let\L=\Lambda 
    \let\Si=\Sigma\let\F=\Phi   
\let\O=\Omega 

\font\tenmib=cmmib10\font\sevenmib=cmmib7\font\fivemib=cmmib5%
\mathchardef\Ba   = "050B  %alpha
\mathchardef\Bb   = "050C  %beta
\mathchardef\Bg   = "050D  %gamma
\mathchardef\Bd   = "050E  %delta
\mathchardef\Be   = "0522  %varepsilon
\mathchardef\Bee  = "050F  %epsilon
\mathchardef\Bz   = "0510  %zeta
\mathchardef\Bh   = "0511  %eta
\mathchardef\Bthh = "0512  %theta
\mathchardef\Bth  = "0523  %vartheta
\mathchardef\Bi   = "0513  %iota
\mathchardef\Bk   = "0514  %kappa
\mathchardef\Bl   = "0515  %lambda
\mathchardef\Bm   = "0516  %mu
\mathchardef\Bn   = "0517  %nu
\mathchardef\Bx   = "0518  %xi
\mathchardef\Bom  = "0530  %omega
\mathchardef\Bp   = "0519  %pi
\mathchardef\Br   = "0525  %rho
\mathchardef\Bro  = "051A  %varrho
\mathchardef\Bs   = "051B  %sigma
\mathchardef\Bsi  = "0526  %varsigma
\mathchardef\Bt   = "051C  %tau
\mathchardef\Bu   = "051D  %upsilon
\mathchardef\Bf   = "0527  %phi
\mathchardef\Bff  = "051E  %varphi
\mathchardef\Bch  = "051F  %chi
\mathchardef\Bps  = "0520  %psi
\mathchardef\Bo   = "0521  %omega
\mathchardef\Bome = "0524  %varomega
\mathchardef\BG   = "0500  %Gamma
\mathchardef\BD   = "0501  %Delta
\mathchardef\BTh  = "0502  %Theta
\mathchardef\BL   = "0503  %Lambda
\mathchardef\BX   = "0504  %Xi
\mathchardef\BP   = "0505  %Pi
\mathchardef\BS   = "0506  %Sigma
\mathchardef\BU   = "0507  %Upsilon
\mathchardef\BF   = "0508  %Phi
\mathchardef\BPs  = "0509  %Psi
\mathchardef\BO   = "050A  %Omega
\mathchardef\BDpr = "0540  %{\bf\partial}
\mathchardef\Bstl = "053F  %{\bf*}

\newdimen\xshift \newdimen\xwidth \newdimen\yshift \newdimen\ywidth

\def\ins#1#2#3{\vbox to0pt{\kern-#2pt\hbox{\kern#1pt #3}\vss}\nointerlineskip}

\def\eqfig#1#2#3#4#5{
\par\xwidth=#1pt \xshift=\hsize \advance\xshift
by-\xwidth \divide\xshift by 2
\yshift=#2pt \divide\yshift by 2
{\hglue\xshift \vbox to #2pt{\vfil
#3 \includegraphics{#4.eps}
}\hfill\raise\yshift\hbox{#5}}}

\def\V#1{{\bf #1}}
\def\lis#1{{\overline#1}}

\font\titolo=cmbx12% scaled\magstep1%
\def\tende#1{\,\vtop{\ialign{##\crcr\rightarrowfill\crcr
 \noalign{\kern-1pt\nointerlineskip} \hskip3.pt${\scriptstyle
 #1}$\hskip3.pt\crcr}}\,}
\def\AA{{\cal A}}\def\DD{{\cal D}}\def\XX{{\cal X}}
\def\EE{{\cal E}}\def\TT{{\cal T}}\def\BB{{\cal B}}\def\LL{{\cal L}}
\def\HH{{\cal H}}\def\NN{{\cal N}}\def\CC{{\cal C}}

\font\msytw=msbm10% scaled\magstep1%
\font\msytww=msbm8 %scaled\magstep1%
\def\RRR{\hbox{\msytw R}}\def\rrr{\hbox{\msytww R}}

\def\ZZZ{\hbox{\msytw Z}}

\def\defi{\,{\buildrel def\over=}\,}
\let\wt=\widetilde

\def\be{\begin{equation}}\def\ee{\end{equation}}
\renewcommand{\theequation}{\arabic{section}.\arabic{equation}}
\let\0=\noindent
\def\*{\vskip2mm}

\def\Eq#1{{\label{#1}}%\write15{\string\Fe{\string#1}{\ref{#1}}}
}
\def\eq#1{(\eqlab{#1})%\write15{\string\Fe{\string#1}{\ref{#1}}}
}
\def\equ#1{(\ref{#1})}
\usepackage{ifthen}
\usepackage{eqalignno}
\usepackage{fancyhdr}
\def\iniz{\setcounter{equation}{0}}
\makeatletter

\newcommand{\Rmnum}[1]{\expandafter\@slowromancap\romannumeral #1@}
\makeatother
\begin{document}

\ifthenelse{\formato=4}{
\centerline{\titolo
Nonequilibrium, thermostats}
\centerline{\titolo  and thermodynamic limit}
{\vskip1mm}

\centerline{G. Gallavotti${}^*$ and E. Presutti${}^@$ }
\centerline{${}^*$ Fisica-INFN Roma1 and Rutgers U.}
\centerline{${}^@$ Matematica Roma2}}
{
\centerline{\titolo
Nonequilibrium, thermostats  and thermodynamic limit}
{\centerline{G. Gallavotti${}^*$ and E. Presutti${}^@$ }
\centerline{${}^*$ Fisica-INFN Roma1 and Rutgers U., ${}^@$ Matematica
  Roma2}}
}

\ifthenelse{\formato=4}{\centerline{\today}}{}

{\vskip1mm}
{\bf Abstract}:
{\it The relation between thermostats of
``isoenergetic'' and ``frictionless'' kind is studied and their
equivalence in the thermodynamic limit is proved in space dimension
$d=1,2$ and, for special geometries, $d=3$.}
\*

\def\Pacs{05.00.00, 05.20.-9, 05.40.-a, 05.70.Ln%
\ifthenelse{\formato=5}{\hfill\today}{}}

\0{pacs: \rm \Pacs}
\*

%%%%%%%%%%%%%%%%%%%%%%%%%%%%%%%%%%%%%%%%%%
\def\SECuno{Introduction}
\def\SECdue{Thermostats}
\def\SECtre{Notations and sizes}
\def\SECquattro{Equivalence: isoenergetic versus frictionless}
\def\SECcinque{Energy bound for frictionless dynamics}
\def\SECsei{Infinite volume. Frictionless dynamics}
\def\SECsette{Entropy bound. Thermostatted dynamics}
\def\SECotto{Conclusions}
\def\AppA{Superstability. Sets of full measures}
\def\AppB{Choice of $R_n(t)$}
\def\AppC{Theorems 2 ($d\le3$) and 3 ($d\le2$)}
\def\AppD{Work bounds. $d=3$ frictionless thermostats}
\def\AppE{Speed bounds. Thermostats in $d=3$}
\def\AppF{Frictionless motion is a flow if $d=2$}
\def\AppG{Free thermostats}
\def\AppH{Quasi invariance}
\def\AppI{Regularized thermostatted dynamics}
\def\AppJ{Entropy bounds: check of Eq.\equ{e7.7}}
\def\AppK{Details on the derivation of Eq.\equ{e7.5},\equ{e7.6}}
\def\AppL{Proof of Lemmas 4,5}
\newcounter{paragrafo}%\setcounter{paragrafo}{27}
\def\piu{{\addtocounter{paragrafo}{1}}\Roman{paragrafo}}
\newcounter{appendice}%\setcounter{paragrafo}{27}
\def\piua{{\addtocounter{appendice}{1}}\Alph{appendice}}

\0{\bf Contents}{\small
\halign{# &\kern3mm#\hfill&\kern3mm#\cr
{\bf\piu}& \SECuno&\pageref{sec1}
\cr
{\bf\piu}& \SECdue&\pageref{sec2}
\cr
{\bf\piu}&  \SECtre&\pageref{sec3}
\cr
{\bf\piu}&  \SECquattro&\pageref{sec4}
\cr
{\bf\piu}&  \SECcinque&\pageref{sec5}
\cr
{\bf\piu}&  \SECsei&\pageref{sec6}
\cr
{\bf\piu}&  \SECsette&\pageref{sec7}
\cr
{\bf\piu}&  \SECotto&\pageref{sec8}
\cr
&  \kern-1.2cm{\it Appendices. Part 1: frictionless dynamics}&
\cr
{\bf\piua}& \AppA&\pageref{A}
\cr
{\bf\piua}& \AppB&\pageref{B}
\cr
{\bf\piua}& \AppC&\pageref{C}
\cr
{\bf\piua}& \AppD&\pageref{D}
\cr
{\bf\piua}& \AppE&\pageref{E}
\cr
{\bf\piua}& \AppF&\pageref{F}
\cr
&  \kern-1.2cm{\it Appendices. Part 2: thermostatted dynamics}&
\cr
{\bf\piua}& \AppG&\pageref{G}
\cr
{\bf\piua}& \AppH&\pageref{H}
\cr
{\bf\piua}& \AppI&\pageref{I}
\cr
{\bf\piua}& \AppJ&\pageref{J}
\cr
{\bf\piua}& \AppK&\pageref{K}
\cr
{\bf\piua}& \AppL&\pageref{L}
%\cr
%{\bf\piua}& \AppM&\pageref{M}
\cr
&  \kern-1.2cm{\it References}&\pageref{ref}
\cr}
}
\makeatletter
\def\iniz{\setcounter{equation}{0}
\ifodd\thepage
{\rhead{\thepage}\lhead{{{{\small\bf\thesection:}\ \SEC}}}}
\else
\lhead{\thepage}\rhead{{\small\bf\thesection:}\ \SEC}
\fi}
\makeatother
%%%%%%%%%%%%%%%%%%%%%%%%%%%%%%%%%%%%%%%%%%
%%%%%%%%%%%%%%%%%%%%%%%%%%%%%%%%%%%%%%%%%%
\def\SEC{\small\SECuno}
\section{\SECuno}\label{sec1}
\iniz
%%%%%%%%%%%%%%%%%%%%%%%%%%%%%%%%%%%%%%%%%%
%%%%%%%%%%%%%%%%%%%%%%%%%%%%%%%%%%%%%%%%%%
In a recent paper \cite{Ga008d} equivalence between isokinetic and
frictionless thermostats has been discussed heuristically, leaving
aside several difficulties on the understanding of the classical
dynamics of systems of infinitely many particles. Such an
understanding is, however, a necessary prerequisite because strict
equivalence can be expected to hold only in the thermodynamic
limit. In this paper we proceed along the same lines, comparing the
isoenergetic and the frictionless thermostats, and study the
conjectures corresponding to the ones formulated in \cite{Ga008d} for
isokinetic thermostats, obtaining a complete proof of equivalence in
$1,2,3$--dimensional systems with various geometries.

In Sec.\ref{sec2} the class of models to which our main result
applies is described in detail. The main result is informally quoted
at the end of the introduction after discussing the physics of the
models; a precise statement will be theorem 1 in Sec.\ref{sec4} and it
will rely on a property that we shall call {\it local dynamics}: the
proof is achieved by showing that in the models considered the
property holds as a consequence of the theorems 2-10, each
of which is interesting on its own right, discussed in the sections
following Sec.\ref{sec4} and in the appendices.  \vskip2mm

A classical model for nonequilibrium statistical mechanics, {\it e.g.}
see \cite{FV963}, is a {\it test system} in a container $\O_0$, and one
or more containers $\O_j$ adjacent to it and enclosing the {\it
interaction systems}.

A geometry that will be considered in dimension $d=2$, to fix the
ideas, can be imagined (see Fig.1, keeping in mind that it is just an
example for convenience of exposition and which could be widely
changed) as in the first picture in Fig.1; the second picture
illustrates the only gemoetry that we shall consider in dimension $d=3$: \*

\eqfig{200}{86}{%
\ins{40}{15}{%
$x=(\V X_0,\dot{\V X}_0,\V X_1,\dot{\V X}_1,\ldots,\V X_\n,\dot{\V X}_\n)$}
}{fig1}{}
\vskip-2mm

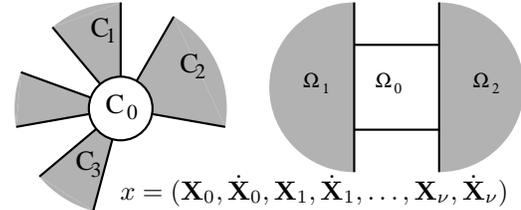
\noindent{Fig.1: \small\it The $1+\n$ finite boxes $\O_j\cap\L,\,
  j=0,\ldots,\n$, are marked $\CC_0,\CC_1,\ldots,\CC_\n$ and contain
  $N_0,N_1,\ldots, N_\n$ particles out of the infinitely many
  particles with positions and velocities denoted $\V X_0,\V
  X_1,\ldots,\V X_\n$, and $\dot{\V X}_0,\dot{\V X}_1,\ldots,$
  $\dot{\V X}_\n$, respecti\-vely, contained in $\O_j,\,j\ge0$. The
  second figure illustrates the special geometry considered for $d=3$:
  here two thermostats, symbolized by the shaded regions, $\O_1,\O_2$
  occupy half-spaces adjacent to $\O_0$. } \*

Referring, for instance, to the first of Fig.1:

\noindent(1) The {\it test system} consists of particles enclosed in a
sphere $\O_0=\Si(D_0)$ of radius $D_0$ centered at the origin.
\vskip2mm

\noindent(2) The {\it interaction systems} consist of particles
enclosed in regions $\O_j$ which are disjoint sectors in $\RRR^d$,
{{\it i.e.}}  disjoint semiinfinite ``spherically truncated'' cones
adjacent to $\O_0$, of opening angle $\o_j$ and axis $\V k_j$:
$\O_j =\{\x\in \RRR^d, |\x|>D_0, \x\cdot\V
k_j>|\x|\,\cos \o_j\}, \,j=1,\ldots,\n$. %%S
\vskip2mm

The initial configurations $x$ of positions and velocities will be
supposed to contain finitely many particles in each unit cube. Thus
the {\it test system} will consist of {\it finitely many
particles}, while the {\it interaction systems are infinitely
exten\-ded}.

We shall suppose that the forces acting on the particles are due to a
repulsive pair interaction of radius $r_\f$ and to external repulsive
interactions acting within a distance $r_\ps$ from the boundaries of
the containers plus in the not frictionless models some {\it
thermostatting forces} (of Gaussian type), \cite{EM990}. Furthermore on
the test system may act a nonconservative ``stirring'' force $\BF$.

The motion starting from $x$ must be defined by first {\it
regularizing} the equations of motion (which are infinitely many and
therefore a ``solution'' has to be shown to exist) ``approximating''
them with evolution equations involving finitely many particles.

There is wide arbitrariness in the choice of the regularization: and
the ambiguity should disappear upon regularization removal.

For instance a first regularization, that we call {\it elastic
regularization}, could be that only the (finitely many) particles of
the initial data $x$ inside an {\it artificial} finite ball
$\L=\Si(r)$ of radius $r>D_0$ will be supposed moving and will be kept
inside $\L$ by an {\it elastic reflection} boundary condition at the
boundary of $\L$.
The particles of $x$ located outside the container $\O_0\cup\cup_{j>0}
(\O_j\cap\L)$ are imagined immobile in the initial positions and
influence the moving particles only through the force that the ones of
them close enough to the boundary of $\L$ exercise on the particles
inside $\L$.

A second {\it alternative} regularization, that we shall call {\it
open regularization}, is obtained by letting only the particles
initially inside $\L$ move while their motion is not influenced by the
particles external to $\L$ and they are even allowed to exit the
region $\L$.

In the ``{\it thermodynamic limit }'', which will be of central
 interest here, the ball $\L$ grows to $\infty$ and the particles that
 eventually become internal to $\L$ start moving: in other words we
 approximate the infinite volume dynamics with a finite volume one,
 called {\it $\L$--regularized}, and then take an infinite volume
 limit.

In the $\L$--regularized evolution the energy in the region
$\Omega_j\cap \Lambda$, $j>0$, will in general change in time.  The
{\it isoenergetic thermostat} is defined by adding ``frictional'' forces
$-\alpha_j \dot q_i$ on all particles $(q_i,\dot q_i)$ in
$\Omega_j\cap \Lambda$ where $\alpha_j$ is chosen in such a way that
the total energy in $\Omega_j\cap \Lambda$ is constant in time, see
the next section for details.  The thermostatted evolution is then the
evolution when such frictional forces are added in each $\Omega_j\cap
\Lambda$, $j>0$.

Isoenergetic models of the kind considered here have
been studied also in simulations aiming at checking the
thermostats``efficiency'', {\it i.e.} the possibility of a boundary
heat exchange sufficient to allow reaching stationarity in systems
with many particles, \cite{GG007}.

The {\it essential physical requirement} that the thermo\-stats should
have a well defined temperature and density will be satisfied by an
appropriate selection of the initial conditions. The guiding idea is
that the thermostats should be so large that the energy that the test
system transfers to them, per unit time in the form of work $Q_j$, is
acquired without changing, at least not in the thermodynamic limit,
the average values of the densities and kinetic energies ({\it i.e.}
temperatures) of the thermostats in any finite observation time
$\Th>0$.

To impose, at least at time $0$ and in the thermodynamic limit, the
requirement the initial data will be sampled with probability $\mu_0$
where $\mu_0$ is the product of extremal DLR distributions in each
$\Omega_j$ with distinct temperatures and chemical potentials, see the
next section for details.
\vskip.5cm

\0{\bf Main result: }{\it In the thermodynamic limit, the
thermostatted evolution, with\-in any prefixed time interval
$[0,\Th]$, becomes identical to the frictionless evolution at least on
a set of configurations which have $\m_0$--probability $1$ with
respect to the initial distribution $\m_0$, in spite of the non
stationarity of the latter. In the same limit also the frictionless
evolution with open or elastic regularization become identical.}  \*
\*

This is proved after theorem 6 in Sec.\ref{sec7}.

The present paper relies heavily on results available in the
literature, \cite{FD977,MPP975,MPPP976,CMP000}, but the results cannot
be directly quoted because (minor) modifications of them are needed
for our purposes. Therefore this paper is self-contained and the
results that could be also found by a careful study of the literature
are proved again from scratch: however we have confined such
derivations in the Appendices A-F with the exception (for the sake of
clarity ot the arguments) of Sec.\ref{sec5},\ref{sec6}; the Appendices
G-L give technical details about the new methods, explained in
Sec.\ref{sec7} and in the Appendices G-J.

The strong restriction on the thermostats geometry in $d=3$ is
commented in Appendix D after the proof of lemma 2, where it is used.

%%%%%%%%%%%%%%%%%%%%%%%%%%%%%%%%%%%%%%%%%%
%%%%%%%%%%%%%%%%%%%%%%%%%%%%%%%%%%%%%%%%%%
\def\SEC{\small\SECdue}
\section{Thermostats}\label{sec2}
%Thermostats
\iniz
%%%%%%%%%%%%%%%%%%%%%%%%%%%%%%%%%%%%%%%%%%
%%%%%%%%%%%%%%%%%%%%%%%%%%%%%%%%%%%%%%%%%%

A configuration $x$ is thus imagined to consist of a configuration
$(\V X_0,\dot {\V X}_0)$ with ${\bf X}_0$ contained in the sphere
$\Si(D_0)$, delimiting the container $\O_0$ of the test systems, and
by $\n$ configurations $(\V X_j,\dot{\V X}_j)$ with $\V X_j\subset
\O_j, \,\dot{\V X}_j\in \RRR^d)$, $j=1,\ldots,\n$:

{\vskip3mm} \noindent {\bf Phase space:} {\it Phase space $\cal H$ is
  the collection of locally finite particle configurations $x=(\ldots,
  q_i,\dot q_i,\ldots)_{i=1}^\infty$
\be x=(\V X_0,\dot{\V X}_0, \V X_1,\dot{\V X}_1, \ldots, \V
X_\n,\dot{\V X}_\n)\defi(\V X,\dot{\V X})
\Eq{e2.1}\ee
with $\V X_j\subset\O_j,\,\dot q_i\in \RRR^d$: in every ball
$\Si(r')$ of radius $r'$ centered at $O$, fall a finite
number of points of $\V X$.}

\* The particles of $x$ located outside $\L$ will be regarded as
immobile or even as non-existing (depending on the choice of elastic
or open regularization considered). {\it It will be convenient to
suppose that the regularization region $\L$ is a ball $\L_n$ of radius
$2^n r_\f$, $n=n_0,n_0+1,\ldots$,} with $n_0$ large so that
$2^{n_0}r_\f>D_0+r_\f$ so that $\L_n$ contain the test system and the
particles interacting with it.

 The particles are supposed to interact with each other, via a
 potential $\f$, and with the non artificial walls ({\it i.e} those of
 the containers $\O_j$), via a potential $\ps$:

\*
\noindent {\bf Interaction:} {\it Interparticle interaction will be
through a pair potential $\f$ with finite range $r_\f$ smooth,
decreasing and positive at the origin. The walls of the containers
$\O_j$ are represented by a smooth decreasing potential $\ps\ge0$ of
range $r_\ps\ll r_\f$ and diverging as an inverse power of the
distance to the walls $\cup_j \partial\O_j$, while the (artificial
boundary) $\partial\L$ will be imagined as a perfectly reflecting
elastic barrier in the case of elastic boundary conditions or as a
boundary perfectly transparent to the moving particles.}

\* Hence the potential $\f$ is {\it superstable} in the sense of
\cite{Ru970}: a property that will play an important role in the
following.  The value of the potential $\f$ at midrange will be
denoted $\lis\f$ and $0<\lis\f\defi\f(\frac{r_\f}2)<\f(0)$; the
wall potential at distance $r$ from a wall will be
supposed given by

\be\ps(r)=\big(\frac{r_\ps}{2r}\big)^{\a}\f_0 ,\qquad r\le
\frac{r_\ps}2
\Eq{e2.2}\ee
with $\a>0$ and $r$ equal to the distance of $q$ to the wall; for
larger $r$ it continues, smoothly decreasing, reaching the value $0$
at $r=r_\ps$.  The choice of $\ps$ as proportional to $\f_0$ limits
the number of dimensional parameters, but it could be made
general. The restriction $r_\ps\ll r_\f$ is not necessary: it has the
physical interpretation of making easier the interaction between
particles in $\O_0$ and particles in $\cup_{j>0} \O_j$ and, therefore,
transfer of energy between test system and thermostats.

Particles in $\O_0$ interact with all the others
but the particles in $\O_j$ interact only with the ones in
$\O_j\cup\O_0$: {\it the test system in $\O_0$ interacts with all
thermostats but each thermostat interacts only with the system}, see
Fig.1.

The $\L$--regularized elastic boundary condition equations of motions
 (see Fig.1), aside from the reflecting boundary condition on the
 artificial boundary of $\L$, concern only the particles in
 $\O_0\cup\cup_{j>0} (\O_j\cap\L)$ and will be

$$\eqalignno{ m\ddot{\V X}_{0i}=&-\partial_i U_0(\V X_0)-\sum_{j>0}
\partial_i U_{0,j}(\V X_0,\V X_j)+\BF_i(\V X_0)\cr m\ddot{\V
X}_{ji}=&-\partial_i U_j(\V X_j)- \partial_i U_{0,j}(\V X_0,\V
X_j)-a\,\a_j \V{{\dot X}}_{ji} &\eq{e2.3}\cr}$$
\\
(1) where the parameter
$a$ will be $a=1$ or $a=0$ depending on the model considered;
\*
\0(2) the potential energies $U_j(\V X_j), \,j\ge0$
and, respectively, $U_{0,j}(\V X_0,\V X_j)$ denote the internal
energies of the various systems and the potential energy of
interaction between the system and the thermostats; hence
for $\V X_j\subset\O_j\cap\L$ the $U_j$'s are:

\ifthenelse{\formato=4}{
\be \eqalign{ U_j(\V X_j)=&\sum_{q\in\V X_j}\ps(q)+\sum_{q,q'\in\V
X_j}\f(q-q')\cr U_{0,j}(\V X_0,\V X_j)=&\sum_{q\in\V
X_0,q'\in\V X_j}\f(q-q');\cr} \Eq{e2.4}\ee}
{\be U_j(\V X_j)=\sum_{q\in\V X_j}\ps(q)+\sum_{q,q'\in\V
X_j}\f(q-q'),\quad U_{0,j}(\V X_0,\V X_j)=\sum_{q\in\V
X_0,q'\in\V X_j}\f(q-q');\Eq{e2.4}\ee}
\0(3) the first label in Eq.\equ{e2.3}, $j=0$ or $j=1,\ldots,\n$,
respectively, refers to the test system or to a thermostat, while the
second labels the points in the corresponding container. Hence the
labels $i$ in the subscripts $(j,i)$ have $N_j$ values and each $i$
corresponds (to simplify the notations) to $d$ components; \*
\0(4) The multipliers $\a_j$ are, for $j=1,\ldots,\n$,

\ifthenelse{\formato=4}{
\be \eqalign{
\a_j{\,{\buildrel def\over=}\,}&\frac{Q_j}{d\,N_j
 k_B T_j(x)/m},\qquad{\rm with}\cr Q_j{\,{\buildrel def\over=}\,}&
-\dot {\V X}_j\cdot \partial_j
U_{0,j}(\V X_0,\V X_j),\cr} \Eq{e2.5} \ee}
{\be \a_j{\,{\buildrel def\over=}\,}\frac{Q_j}{d\,N_j
 k_B T_j(x)/m},\qquad{\rm with},\qquad Q_j{\,{\buildrel def\over=}\,}
-\dot {\V X}_j\cdot \partial_j
U_{0,j}(\V X_0,\V X_j), \Eq{e2.5} \ee}
where $\frac{d}2\, N_j\,k_B\, T_j(x)\defi K_{j,\L}(\dot{\V X}_j)\defi
\frac{m}2\dot{\V X}_j^2$ and $\a_j$ are chosen so that
$K_{j,\L}(\dot{\V X}_j)+U_{j,\L}(\V X_j)=E_{j,\L}$ are {\it exact
constants of motion}: the subscript $\L$ will be omitted unless really
necessary. A more general model to which the analysis that follows
also applies is in \cite{Ga006c}.
\* \0(5) The forces $\BF(\V X_0)$ are, positional, {\it
nonconservative}, smooth ``stirring for\-ces'' (possibly absent).  \*
\0(6)\label{t} In the case of $\L$-regularized thermostatted dynamics
we shall consider only initial data $x$ for which the kinetic energies
$K_{j,\L}(\dot{\V X}_j)$ of the particles in the $\O_j\cap\L$'s are
$>0$ for all large enough $\L$. Then the time evolution is well
defined for $t\le t_\L(x)$ where $t_\L(x)$ is the maximum time before
which the kinetic energies remain positive (hence the equations of
motion remain defined for $t<t_\L(x)$ because the denominators in the
$\a_j$ stay $>0$; see Appendix I for technical details).  \*

The equations with $a=1$ will be called $\L$-re\-gularized
{\it isoenergetically thermostatted} because the energies
$E_j=K_j+U_j$ stay exactly constant for $j>0$ and equal to their
initial values $E_j$.  The equations with $a=0$ in Eq.\equ{e2.3} will
be considered together with the above and called the $\L$--regularized
{\it frictionless equations}.
\*

{\it Remark that $Q_j$ is the work done, per unit time, by the test
system on the particles in the $j$-th thermostat: it will therefore be
interpreted as heat ceded to the $j$--th thermostat. }  \*

\*
It will be convenient to consider also the open boundary condition at
least in the frictionless thermostats case: the equations of
motion are immediately written.

To impose, at least at time $0$ and in the thermodynamic limit, the
requirement that the thermo\-stats should have a well defined
temperature and density the values $N_j, E_j,\, j>0$, will be such
that $\frac{N_j}{|\O_j\cap\L|}$ $\tende{\L\to\infty} \d_j$ and
$\frac{E_j}{|\O_j\cap\L|}\tende{\L\to\infty} e_j$: with $\d_j,e_j >0$
fixed in a sense that is specified by a choice of the initial data
that will be studied, and whose physical meaning is that of imposing
the values of density and temperature in the thermostats, for $j>0$.

{\vskip3mm} \noindent {\bf Initial data:} {\it The probability
  distribution $\m_0$ for the random choice of initial data will be,
  if $dx{\,{\buildrel def\over=}\,}\prod_{j=0}^\n\frac{ d\V
  X_j\,d\dot{\V X}_j}{N_j!}$, the limit as $\L_0\to\infty$ of the
  \ifthenelse{\formato=4}{ finite volume grand canonical distributions
  on ${\cal H}$

$$\eqalignno{
&\m_{0,\L_0}(dx)=const\,\, e^{-H_{0,\L_0}(x)}
\,dx,\qquad{\rm with}&{\rm\eq{e2.6}}\cr
&H_{0,\L_0}(x)\defi \sum_{j=0}^\n \b_j (K_{j,\L_0}
(x)-\l_j N_{j,\L_0}+ U_{j,\L_0}(x))\cr
&\b_j{\,{\buildrel def\over=}\,} \frac1{k_BT_j}>0,\,\l_j\in \RRR,
\cr}$$}
{finite volume grand canonical distributions on ${\cal H}$
$\m_{0,\L_0}(dx)=const\,\, e^{-H_{0,\L_0}(x)}
\,dx$, with
\be
H_{0,\L_0}(x)\defi \sum_{j=0}^\n \b_j (K_{j,\L_0}
(x)-\l_j N_{j,\L_0}+ U_{j,\L_0}(x)),\qquad
\b_j{\,{\buildrel def\over=}\,} \frac1{k_BT_j}>0,\,\l_j\in \RRR,
\Eq{e2.6}\ee}
}

\noindent{\it Remarks:} (a) The values $ \b_0=\frac1{k_BT_0}>0,\l_0\in\RRR$,
are also fixed, although they bear no particular physical meaning because the
test system is kept finite.
\\
(b) Here $\Bl=(\l_0,\l_1,\ldots\l_\n)$ and ${\bf T}=(T_0,T_1,\ldots, T_\n)$
are fixed {\it chemical potentials} and {\it temperatures}, and $\L_0$
is a ball centered at the origin and of radius $D_0$.
\\ (c) The distribution $\m_0$ is a Gibbs distribution obtained by
taking the ``thermodynamic limit'' $\L_0\to\infty$. The theory of the
thermodynamic limit implies the existence of the limit distribution
$\m_0$, either at low density and high temperature or on subsequences,
\cite{Ga000}. In the second case (occurring when there are phase
transitions at the chosen values of the thermostats parameters)
boundary conditions have to be imposed that imply that the thermostats
are in a pure phase: for simplicity such exceptional cases will not be
considered; this will be referred to as a ``no-phase
transitions''\label{npt} restriction.
\\ (d) $\L_0$ should not be confused with the regularization sphere
$\L$: it is introduced here and made, right away, $\infty$ only to
define $\m_0$.
\\ (e) Notice that $\m_0$ is a product of {\it independent equilibrium
Gibbs distributions} because $H_0$ does not contain the interaction
potentials $U_{0,j}$.
\\ (f) The proofs extend to $\m_0$'s which are product of DLR
distributions in each container (which are not necessarily extremal).

\*

%%%%%%%%%%%%%%%%%%%%%%%%%%%%%%%%%%%%%%%%%%
%%%%%%%%%%%%%%%%%%%%%%%%%%%%%%%%%%%%%%%%%%
\def\SEC{\small\SECtre}
\section{\SECtre}\label{sec3}
%Notations and sizes
\iniz
%%%%%%%%%%%%%%%%%%%%%%%%%%%%%%%%%%%%%%%%%%
%%%%%%%%%%%%%%%%%%%%%%%%%%%%%%%%%%%%%%%%%%

Initial data will be naturally chosen at random with
respect to $\m_0$.  Let the ``pressure'' in the $j$-th thermostat be
defined by $p_j(\b,\l;\L_0) {\buildrel def\over=}
\frac1{\b\,|\O_j\cap\L_0|}$ $\log Z_{j,\L_0}(\b,\l)$ with

\be
Z_{\L_0}(\b,\l)=\sum_{N=0}^\infty \int \frac{dx_N}{N!}
e^{-\b(-\l
  N+K_{j}(x_N)+U_{j}(x_N))}
\Eq{e3.1}\ee
where the integration is over positions and velocities of the particles
in $\L_0\cap\O_j$. Defining $p(\b,\l)$ as the thermodynamic limit,
$\L_0\to\infty$, of $p_j(\b,\l;\L_0)$ we shall say that the
thermostats have densities $\d_j$, temperatures $T_j$, energy
densities $e_j$ and potential energy densities $u_j$, for $j>0$, given
by equilibrium thermodynamics, {\it i.e.}:

$$\eqalignno{
\d_j=&-\frac{\partial p( \b_j,\l_j)}{\partial\l_j},\quad
k_BT_j= \b_j^{-1}&\eq{e3.2}\cr
e_j=&-\frac{\partial \b_jp(\b_j,\l_j)}{\partial \b_j}-\l_j\d_j,
\qquad u_j=e_j-\frac{d}2 \d_j\b_j^{-1}\cr}$$
which are the relations linking density $\d_j$, temperature
$T_j=(k_B\b_j)^{-1}$, energy density $e_j$ and potential energy
density $u_j$ in a grand canonical ensemble.

In general the $\L$--regularized time evolution changes the
$\m_0$-measure of a volume element in phase space by an amount related
to (but different from) the variation of the Liouville volume because,
in general,

\be
\s(x)=\frac{d}{dt}\log\frac{\m_0(S_tdx)}{\m_0(dx)}\big|_{t=0}\not\equiv
0\Eq{e3.3}\ee
if $x\to S_tx$ denotes the solution of the equations of motion (for the
considered model). The variation $\s(x)$, per unit time of a volume
element, in the sector of phase space containing $N_j>0$ particles in
$\O_j\cap\L,\, j=0,1,\ldots,\n$, can be computed and is, under the
$\L_n$--regularized dynamics and for the elastic boundary conditions models,

$$\eqalignno{
\s(x)=&\sum_{j\ge0}\b_j Q_j,\kern2cm a=0&\eq{e3.4} \cr
\s(x)=&\sum_{j>0}\frac{Q_j}{k_B T_j(x)}\,{(1-\frac1{d\,N_j})}+
\b_0 Q_0,\qquad a=1\cr
}$$
as it follows by adding the time derivative $\b_0 Q_0 \defi$
$\b_0(\dot K_0+\dot U_0)$ to the divergence of Eq.\equ{e2.3} (regarded as a
first order equation for the $q$'s and $\dot q$'s) using the
expression in Eq.\equ{e2.5} for $\a_j$.  \*

\noindent {\it Remarks:} (1) The expressions for $\s(x)$ will play a
central role in our approach and one can say that the key idea of this
work is to control the size of $|\s|$ and through it show that
``singular'' events like all particles come close to a simultaneous
stop or accumulate to a high density somewhere or some of them acquire
very high speed or approach too closely the walls (hence acquiring
huge potential energy) will have zero $\m_0$-probability. Such events
have zero $\m_0$-probability at time $0$ and the variations with time
of $\m_0$ are controlled by $|\s|$. Naturally, in view of
Eq.\equ{e3.4}, we shall call such estimates ``{\it entropy bounds}.

\0(2) The relation $\b_0\,(\dot K_0+\dot
  U_0)=\b_0\,(\BF\cdot \dot{\V X}_0-\sum_{j>0}(\dot U_{0j}-Q_j))$ is
  useful in studying Onsager reciprocity and Green-Kubo formulae,
  \cite{Ga008c}. Notice that $\dot K_0+\dot U_0\defi Q_0$ is also
  $-\sum_{j>0} \dot{\V X}_0\cdot\BDpr_{\V
  X_0}U_{0,j}(\V X_0,\V X_j)+ \BF\cdot \dot{\V X}_0$.

\0(3) It is also interesting to consider {\it isokinetic}
thermostats: the multipliers $\a_j$ are then so defined that $K_j(x)$ is
an exact constant of motion: calling its value $\frac32 N_jk_B T_j(x)$
the multiplier $\a_j$ becomes

\be \a_j(x)\,{\,{\buildrel def\over=}\,}\frac{Q_j- \dot U_j}{d\,N_j
 k_B T_j(x)/m}, \Eq{e3.5} \ee
with $Q_j$ defined as in Eq.\equ{e2.5}. They have been studied
heuristically in \cite{Ga008d}.
\*

The open boundary conditions will only be considered for the
$\L_n$--regularized frictionless dynamics ($\L_n$ is defined in the
paragraph following Eq.\equ{e2.1}) and denoted $x\to \lis
S^{(n,0)}_tx$: in this case only the particles initially inside $\L_n$
will move freely allowed to cross $\partial\L_n$ unaffected by the
particles initially outside $\L_n$ which will keep their positions
fixed.

The $\L_n$--regularized motions with elastic boundary conditions will
be denoted $x\to S^{(n,a)}_tx,\, a=0,1$.

Let $S^{(a)}_tx$ or $\lis S^{(0)}_tx$ be the infinite volume dynamics
$\lim_{n\to\infty}$ $ S_t^{(n,a)}x$, $a=0,1$, or
$\lim_{n\to\infty}\lis S^{(n,0)}_tx$ {\it provided the limits
exist}. We shall often use the notations

\be\eqalign{
x^{(n,a)}(t)\defi& S^{(n,a)}_t x,\qquad
\lis x^{(n,0)}(t)\defi \lis S^{(n,0)}_t x,\cr
x^{(0)}(t)\defi& S^{(0)}_t x,\qquad\kern5mm \lis x^{(0)}(t)\defi
\lis S^{(0)}_t x.\cr}\Eq{e3.6}\ee

\0{\it Remarks:} (1) In the frictionless case the existence of a
solution to the equations of motion poses a problem only if we wish to
study the $\L_n\to\infty$ limit, {{\it i.e.\ }} in the case in which the
ther\-mostats are infinite: for $\L_n$ finite $\lis S^{(n,0)}_tx$ is
well defined. While for the elastic reflections at the artificial
regularization boundary $\partial\L_n$
it is shown in \cite{MPPP976}  that the dynamics is also well
defined with $\m_0$-probability $1$.  \\
(2) In the thermostatted case the kinetic energy appearing in the
denominator of $\a_j$, see Eq.\equ{e2.5}, can be supposed to be $>0$
with $\m_0$--probability $1$ at $t=0$. However it can become $0$ later
at $t_{\L_n}(x)$ (see item (6), p.\pageref{t}, and the example at end
of Sec.\ref{sec4}). In the course of the analysis it will be proved
that with $\m_0$--probability $1$ it is
$t_{\L_n}(x)\tende{n\to\infty}\infty$; therefore $S^{(n,1)}_tx$ is
eventually well defined.  \\
 (3) It will be shown that the limits
$\lis x^{(0)}(t)$ and $x^{(0)}(t)$ exist and are identical (as
expected). It will become clear why it is necessary to consider first
the frictionless thermostats with open boundary conditions.  \*

We shall denote $(S_t^{(n,a)}x)_j$ or $x^{(n,a)}_j(t)$ the
positions and velocities of the particles of $S^{(n,a)}_t x$ in
$\O_j$; a corresponding notation will be used for positions and
velocities of $\lis S^{(n,0)}_t x$ .

If $x^{(n,a)}_{ji}(t)$ denote the pairs of positions and velocities
$(q^{(n,a)}_i(t),\dot q_i^{(n,a)}(t))$ with $q_i\in\O_j$.  Then a
particle with coordinates $(q_i,\dot q_i)$ at $t=0$ in, {\it say}, the
$j$-th container evolves {\it between collisions} with the regularization
boundary $\L_n$ (if any), see Eq.\equ{e2.3}, as

\ifthenelse{\formato=4}{
\kern-6mm
\be\eqalign{
q_i(t)=&q_i(0)+\int_0^t\dot q_i(t')\,dt'\cr
\dot q_i(t)=&e^{-\int_{0}^ta\,\a_j(t')dt'}\dot q_i(0)\cr
&+\int_{0}^t
e^{-\int_{t''}^t a\,\a_j(t')dt'} {f_i(x^{(n,a)}(t''))}
\,dt''\cr}\Eq{e3.7}\ee}
{\be\eqalign{
q_i(t)=&q_i(0)+\int_0^t\dot q_i(t')\,dt'\cr
\dot q_i(t)=&e^{-\int_{0}^ta\,\a_j(t')dt'}\dot q_i(0)+\int_{0}^t
e^{-\int_{t''}^t a\,\a_j(t')dt'} {f_i(x^{(n,a)}(t''))}
\,dt''\cr}\Eq{e3.7}\ee}
where $m f_i=-\partial_{q_i} \big(U_j(\V X_j(t))+U_{0,j}(\V X_0(t),{\V
X}_j(t))\big)+\d_{j0}\F_i(\V X_0(t))$ and $\V X_j(t)$ denotes $\V
X_j^{(n,a)}(t)$ or $\lis{\V X}^{(n,0)}_j(t)$. Here $U_j$ is defined as
in Eq.\equ{e2.4}; $\a_0\equiv0$.  \*

{The open boundary evolution is described by Eq.\equ{e3.7} with
  $a=0$ and with the appropriate interpretation of $f_i$.}  \*

The first difficulty with infinite dynamics is to show that the number
of particles, and their speeds, in a finite region $\L$ remains finite and
bounded only in terms of the region diameter $r$ (and of the initial data):
for all times or, at least, for any prefixed time interval.

We shall work with dimensionless quantities: therefore
suitable choices of the units will be made. If $\Th$ is prefixed
as the maximum time that will be considered, then

\be\eqalign{
&\f_0\,: \, \hbox{(energy scale)},\
r_\f\,: \, \hbox{(length scale)},\cr
&\Th \,: \, \hbox{(time scale)},\
v_1=\sqrt{\frac{2\f(0)}m}\,
\ \hbox{(velocity scale)}\cr}\Eq{e3.8}\ee
are natural units for energy, length, time, velocity.

It will be necessary to estimate quantitatively the size of various
kinds of energies of the particles, of a configuration $x$, which are
localized in a region $\D$. Therefore introduce $e(\dot q,q)\defi
\big(\frac{{m\dot q}^2}{2} +
\psi(q)\big)/\f_0$ and, for any region $\D$,
the following dimensionless quantities:

\be\eqalign{
(a)\ &N_\D(x),N_{j,\D}(x)\ \hbox{the number of particles of $x$}\cr
& \hbox{located in $\D/\O_0$ or, respectively, $\D\cap\O_j$}\cr
(b)\ &e_\D(x)\defi\,\max_{q_i\in\D/\O_0}\,e(\dot q_i,q_i)
\cr
(c)\ &U_\D(x)=\frac1{2} \sum_{q_i,q_j\in\D/\O_0,\, i\ne j}
\f(q_i-q_j)/\f_0\cr
(d)\ &V_\D(x)= \max_{q_i\in\D/\O_0} {|\dot q_i|}/{v_1}
\cr
}\Eq{e3.9}\ee

The symbol ${\BB}(\x,R)$ will denote the ball centered at $\x$ and
with radius $R \,r_\f$. With the above notations the {\it local
dimensionless energy} of the thermostat particles in ${\cal B}(\x,R)$
will be defined as
% $W(x;\x,R) \defi$ $ \e_{\BB(\x,R)}(x)+
%$$U_{\BB(\x,R)}(x)+N_{\BB(\x,R)}(x)$ or, more explicitly,

\ifthenelse{\formato=4}{\be
\eqalign{ &W(x;\x,R) \defi\frac1{\f(0)}\sum_{q_i\in {\cal
B}(\x,R)/\O_0}\Big(\frac{m\dot q_i^2}2+\ps(q_i)\cr
&+\frac12\sum_{q_i,q_j\in\BB(\x,R),\, i\ne j}
\f(q_i-q_j)+\f(0) \Big)\cr}\Eq{e3.10}\ee}
{\be
W(x;\x,R) \defi\frac1{\f(0)}\sum_{q_i\in {\cal
B}(\x,R)/\O_0}\Big(\frac{m\dot q_i^2}2+\ps(q_i)
+\frac12\sum_{q_j\in\BB(\x,R)/\O_0,\, i\ne j}
\f(q_i-q_j)+\f(0) \Big)\Eq{e3.10}\ee}
Let $\log_+ z\defi\max\{1,\log_2|z|\}$, $g_\z(z)=(\log_+ z)^\z$ and

\be
\EE_\z(x)\defi \sup_{\x}\sup_{R> g_\z(\frac{\x}{r_\f})} \frac{W(x;\x,R) }{R^d}
\Eq{e3.11}\ee

If $\HH$ is the space of the locally finite configurations ({\it i.e.}
containing finitely many particles in any finite region), let
${\cal H}_\z\subset\HH$ be the configurations with

\be\eqalign{
(1) \ &\ {\cal  E}_\z(x)<\infty,\qquad
(2) \ \frac{K_{j,\L}}{|\L\cap\O_j|}\,
>\frac12\frac{\d_j\,d}{2\b_j}\cr}\Eq{e3.12}\ee
for all $\L={\cal B}(O,L)$ large enough and for $\d_j,T_j$, given by
Eq.\equ{e3.2}; here $N_{j,\L}$, $U_{j,\L}, K_{j,\L}$ denote the number
of particles and their potential or kinetic energy in
$\O_j\cap\L$.
\*

\0{\it Remark:} Notice that the lower bound in (2) of Eq.\equ{e3.12}
is {\it half the value of the average kinetic energy} in the initial
data and it will become clear that any prefixed fraction of the
average kinetic energy could replace $1/2$.  Each set ${\cal H}_\z$
has $\m_0$-probability $1$ for $\z\ge1/d$, see Appendix A.

%%%%%%%%%%%%%%%%%%%%%%%%%%%%%%%%%%%%%%%%%%
%%%%%%%%%%%%%%%%%%%%%%%%%%%%%%%%%%%%%%%%%%
\def\SEC{\small Equivalence}
\section{\SECquattro}\label{sec4}
%Equivalence: isoenergetic versus frictionless
\iniz
%%%%%%%%%%%%%%%%%%%%%%%%%%%%%%%%%%%%%%%%%%
%%%%%%%%%%%%%%%%%%%%%%%%%%%%%%%%%%%%%%%%%%

Adapting a conjecture, in \cite{Ga008d}, we are led
to expect that the motions considered satisfy the property:

\vskip3mm
\noindent {\bf Local dynamics} {\it Given $\Th>0$, for $t\in [0,\Th]$
  and with $\m_0$--probability $1$:
\\
(1) The limits
$x^{(a)}(t)\defi $ $\mathop{\lim}\limits_{n \to\infty}$ $
x^{(n,a)}(t)$, $\lis x^{(0)}(t)\defi\mathop{\lim}\limits_{n\to\infty}$ $
\lis x^{(n,0)}(t)$
(``thermo\-dyna\-mic limits'') exist for all $t\le \Th$ and $a=0,1$.
\\ (2) For $t\le\Th$, $x^{(n,1)}(t)$ satisfies the second of
Eq.\equ{e3.12}.
\\
(3) The functions $t\to x^{(0)}(t)$ and $t\to \lis x^{(0)}(t)$ solve
uniquely the frictionless equations in a subspace of $\HH$ to which
also $x^{(1)}(t)$ belongs (explicit, sufficient, bounds are described
in theorem 5).}
\*

\noindent{\it Remarks:} (a) The limits of $x^{(n,a)}(t), \lis
x^{(n,0)}(t)$, as $\L_n\to\infty$, are understood in the sense that for
any ball $\D$ whose boundary does not contain a particle of
$x^{(0)}(t)$ the labels of the particles of $x^{(a)}(t), \lis
x^{(0)}(t)$ and those of the particles in $x^{(n,a)}(t),\lis
x^{(n,0)}(t)$ which are in $\D$ are the same, and for each $i$ the
limits of $ (q_i^{(n,a)}(t), \dot q_i^{(n,a)}(t))$ and $ (\lis
q_i^{(n,a)}(t), \dot {\lis q}_i^{(n,a)}(t))$ exist and are
continuous, together with their first two derivatives.% for each $i$.
\*

\noindent(b) Uniqueness in item (3) can be given several meanings. The
simplest is to require uniqueness in the spaces $\HH_\z$ for
$\z\ge1/d$ fixed: and theorem 11, in Appendix F, shows that for $d=1,2$
one could suppose such simpler property. However our statement is more
general and we have left deliberately undetermined which subspace is
meant in item (3) so that the determination of the subspace has to be
considered part of the problem of establishing a local dynamics
property. The generality becomes relevant in studying the case $d=3$,
where even in equilibrium a proof that the evolution of data in
$\HH_\z$ remains in the same space is lacking. The local dynamics
property in $d=3$ is implied by theorem 9, in Appendix E.
\*

\noindent(c) Recalling the characteristic velocity scale (namely
$v_1=\sqrt{{2\f(0)}/{m}}$), the initial speed of a particle located in
$q\in\RRR^d$, is bounded by $v_1\sqrt{\EE_{1/d}}
g_{1/d}(q/r_\f)^{{d}/2}$; and the distance to the walls
of the particle located at $q$ is bounded by $({\EE_{1/d}}
g_{1/d}(q/r_\ps)^d)^{-1/\a}\,r_\f $.

Hence for $|q|$ large they are, respectively, bounded
proportionally to $[(\log |q|/r_\f)^\frac1d]^{\frac{d}2}$ and $[(\log
|q|/r_\f)^\frac1d]^{-\frac{d}\a}$: this says that locally the particles
have, initially, a finite density and reasonable energies and velocity
distributions (if measured on boxes of a ``logarithmic scale''). The
theorem 11 in Appendix F will show that this property remains true for
all times, with $\m_0$ probability $1$.
\*

\noindent(d) An implication is that Eq.\equ{e3.7} has a meaning at
time $t=0$ with $\m_0$--probability $1$ on the choice of the initial
data $x$, because $\EE(x)<\infty$.  \*

\noindent(e) The further property that the thermostats are {\it
efficient}: {\it i.e.} the work performed by the external non
conservative forces is actually absorbed by the thermostats in the
form of heat $Q_j$, so that the system can eventually reach a
stationary state, will not be needed because in a finite time the
external forces can only perform a finite work (if the dynamics is
local).  \*

\noindent (f) It should also be expected that, with
$\m_0$--probability $1$, the limits as $\L\to\infty$ of item (2),
Eq.\equ{e3.12}, should exist and be equal to $\frac{d\,\d_j}{2\b_j}$
for almost all $t\ge0$ respectively: this is a question left open (and
it is not needed for our purposes).  \*

{\it Assuming the local dynamics} property, equivalence, hence the
property $\lis x^{(0)}(t)\equiv x^{(1)}(t)\equiv x^{(0)}(t)$ for all
finite $t$, can be established as in \cite{Ga008d}. This is recalled
checking, for instance, $\lis x^{(0)}(t)\equiv x^{(1)}(t)$, in the
next few lines.

In the thermostatted $\L_n$--regularized case it is
\be |\a_j(x)|=\frac{|\dot{\V X}_j\cdot\partial_j U_{0,j}(\V
X_0,\V X_j)|}{
\dot {\V X}_j^2}\Eq{e4.1} \ee
The force between pairs located in $\O_0,\O_j$ is bounded by
$F{\,{\buildrel def\over=}\,} \max|\partial \f(q)|$; the numerator of
Eq.\equ{e4.1} can then be bounded by $F N_0 \sqrt{\lis
N_\Th}\sqrt{2K_j/m}$ where $N_0$ is the number of particles in $\CC_0$
and $\lis N_\Th$ bounds the number of thermostat particles that can be
inside the shell of radii $D_0$, $D_0+r_\f$ for $0\le t\le \Th$
(by Schwartz' inequality).

Remark that the bound on $\lis N_\Th$ exists by the local evolution
hypothesis (see (1) and remark (a)) but, of course, is not uniform
in the initial data $x$. Hence for $0\le t\le\Th$ and large enough
$n$ Eq.\equ{e3.12} yields:

\be |\a_j|\le \frac{\sqrt{m}F N_0\sqrt{\lis N_\Th}}{\sqrt{2
    K_{j,\L_n}(x^{(1, \L_n)}(t))}}\le\frac{\sqrt{m}F N_0\sqrt{\lis
    N_\Th}}{\sqrt{|\O_j\cap \L_n| \d_j d/2\b_j}}.\Eq{e4.2} \ee
Letting
$\L_n\to\infty$ it follows that $\a_j\tende{\L_n\to\infty}0$.

Taking the limit $\L_n\to\infty$ of Eq.\equ{e3.7} {\it at fixed $i$},
this means that, {\it with $\m_0$--probability $1$}, the limit motion as
$\L_n\to\infty$ (with $\b_j, \l_j,\,j>0, $ constant) satisfies

\be
q_i(t)=q_i+\int_0^t \dot q_i(t')dt',\, \dot q_i(t)=\dot q_i+\int_0^t
f_i(t'') dt''\Eq{e4.3}\ee
{\it i.e.} the frictionless equations; and the solution
to such equations is unique with probability $1$ (having again used
assumption (3) of the local dynamics). In conclusion\*

\0{\bf Theorem 1: }{\it If the dynamics is local in the above sense
then in the thermodynamic limit the thermostatted evolution, within
any prefixed time interval $[0,\Th]$, becomes the frictionless
evolution at least on a set of configurations which have probability
$1$ with respect to the initial distribution $\m_0$, in spite of the
non stationarity of the latter.}
\*

Suppose, in other words, that the initial data are sampled with the
Gibbs distributions for the thermostat particles (with given
temperatures and densities) and with an {\it arbitrary distribution}
for the finite system in $\O_0$ with density with respect to the
Liouville volume (for instance with a Gibbs distribution at
temperature $(k_B\b_0)^{-1}$ and chemical potential $\l_0$ as in
Eq.\equ{e2.6}).

Then, {\it in the thermodynamic limit}, the time evolution is the same
that would be obtained, in the same limit $\L_n\to\infty$, via a
isoenergetic thermostat acting in each container $\O_j\cap\L_n$ and keeping
its total energy (in the sector with $N_j$ particles) constant and with
a density equal (asymptotically as $\L_n\to\infty$) to $e_j$.

The difficulty of proving the locality property (2) cannot be
underestimated, although it might seem, at first sight, ``physically
obvious'': the danger is that evolution implies that the thermostat
particles {\it grind to a stop} in a finite time converting the
kinetic energy entirely into potential energy. The consequence would
be that $\a_j$ becomes infinite and the equations ill defined.

As a consequence it is natural to expect, as stated in the local
dynamics assumption, only a result in $\m_0$--probability. This can be
better appreciated considering the following {\it counterexample}, in the
frictionless case.

Consider an initial configuration in which particles are on a square
lattice (adapted to the geometry): regard the lattice as a set of
adjacent tiles {\it with no common points}. Imagine that the particles
at the four corners of each tile have velocities of equal magnitude
pointing at the center of the tile. Suppose that the tiles sides are
$> r_\f$. If $\f(0)$ is large enough all particles come to a stop in
the same finite time and at that moment all kinetic energy has been
converted into potential energy: at time $0$ all energy is kinetic and
later all of it is potential. Certainly this example, which concerns a
single event that has, therefore, $0$ probability in $\m_0$, shows
that some refined analysis is necessary: the thermostatted evolution
$x^{(n,1)}(t)$ might be not even well defined because the denominator
in the definition of $\a_j$ might become $0$.  \*

It should be stressed that the thermostats models considered here
preserve, even at finite $\L_n$, an important symmetry of nature: {\it
time reversal}: this certainly explains the favor that they have
received in recent years in the simulations.

A corollary will be that neither the frictionless motion nor
the dissipative thermostatted motions conserve phase space volume
(measured with $\m_0$), but in both cases the entropy production rate
coincides with the phase space volume (measured by $\m_0$) contraction
and, at the thermodynamic limit, is identical in the two cases.

In Appendix G we have discussed the very simple case in which there is
no direct mutual interaction between the thermostats particles, {\it
  free thermostats}, and the walls are reflecting. Reading this simple
case may help understanding the organization of the analysis in the
interacting cases.

%%%%%%%%%%%%%%%%%%%%%%%%%%%%%%%%%%%%%%%%%%%
%%%%%%%%%%%%%%%%%%%%%%%%%%%%%%%%%%%%%%%%%%%
\def\SEC{\small\SECcinque}
\section{\SECcinque}\label{sec5}
\iniz
%\def\SECcinque{Energy bound for frictionless dynamics}
%%%%%%%%%%%%%%%%%%%%%%%%%%%%%%%%%%%%%%%%%%%
%%%%%%%%%%%%%%%%%%%%%%%%%%%%%%%%%%%%%%%%%%%

Proof of the local dynamics property will require controlling the
maximal particles speeds, the number of particles interacting with any
given one as well as their number in any finite region. This will be
achieved by proving bounds on the local energies $W(x;\x,R)$,
Eq.\equ{e3.10}. For instance the speed $|\dot q|$ of a particle
$q,\dot q$ in $x$ is bounded by ${\dot q}^2\le v_1^2 W(x,q,R)$, $R>0$,
and its distance $\rho$ from the walls by $\rho \ge \frac{r_\ps}2
W(x,q,R)^{-\frac 1\alpha}$, $R>0$.

In the geometries described in Fig.1 we shall preliminarily discuss
bounds at time $0$, {\it e.g} Eq.\equ{e5.3} below, and then we shall
use energy conservation to extend the bounds to positive time.

Superstability of the potential $\f$ implies that the number $N$ of
points in a region $\D$, a cube or a ball, can be bounded in terms of
the potential energy $U$ in the same region and of
$\f_0=\f(0),\lis\f=\f(\frac{r_\f}2)>0$ (defined after
Eq.\equ{e2.1}). The bound derivation is recalled in Appendix A for
clarity, and yields the inequality

\be N_\D\le C \frac{\sqrt W}{\sqrt{|\D|}},\qquad C\defi
\big(\frac{2\f_0}{\lis\f}\big)^{\frac12},\Eq{e5.1}\ee

Calling $\EE\defi\EE_{1/d}(x)$, Eq.\equ{e3.11}, consider the sequence
of balls $\L_n={\cal B}(O,2^n)$, of radii $L_n= 2^n r_\f$, $n\ge n_0$,
see paragraph following Eq.\equ{e2.1}.  Given a configuration $x$ let
$N(x;\x,R)$ the number of particles in the ball of radius $R\,r_\f$
centered at $\x$ and

\be\eqalign{
(1)\,\, &  V_n = \,(\hbox{ max velocity in}\ \L_n/\O_0\,)/v_1\cr
(2)\,\, &  \r_n= \,\hbox{ min distance to }\,
\partial (\O_j\cap\L_n) \, {\rm of}\, q_i\not\in\O_0\cr
(3)\,\, &  \NN_n=\, \max_{q_i\in\L_n/\O_0} N(x;q_i,1)\cr}\Eq{e5.2}\ee
Such quantities can be bounded in terms of the
maximum of $W(x,\xi,R)$ over $\xi \in \Lambda_n$ and, by choosing
$R=n^{1/d}$, we obtain, see definitions Eq.\equ{e3.8},

\be \kern-2mm V_n=%v_1
\, (n\EE)^{\frac12},\, \r_n = {(n\,
{\EE})^{-\frac1\a}},\, \NN_n\le C(n\EE)^{\frac12} \Eq{e5.3}\ee
under the assumption that the wall potential has range $r_\ps$ and is
given by Eq.\equ{e2.2}; the last inequality is a consequence of the
definition of $W$ and of the above mentioned superstability,
Eq.\equ{e5.1}. By controlling the growth in time of the energies $W$
we shall extend the validity of Eq.\equ{e5.3} to positive times.  \*

\0{\it Constants convention:} From now on we shall encounter various
constants that are all computable in terms of the data of the problem
(geometry, mass, potentials, densities, temperatures and the
(arbitrarily) prefixed time $\Th$): Eq.\equ{e5.1} gives a simple
example of a computation of a constant. To avoid proliferation of
labels all constants will be positive and denoted $C,C',C'',\ldots,
B,B',\ldots$ or $c,c',c'', \ldots, b,b',b'',\ldots$: they have to be
regarded as functions of the order of appearance, non decreasing the
ones denoted by capital letters and non increasing the ones with lower
case letters; furthermore the constants $C,\ldots,c,\ldots$ may also
depend on the parameters that we shall name $\EE$ or $E$ and will be
again monotonic non decreasing or non increasing, respectively, as
functions of the order of appearance and of $\EE$ or $E$.  \*

{\it In this and in the next section we shall study only the open
  regularization $\lis x^{(n,0)}(t)$ $\defi\lis S^{(n,0)}_tx$, $t\le
  \Th$, $\Th$ being an arbitrarily fixed positive time.}  Therefore
  $\lis x^{(n,0)}(\t)$ will be a finite configuration of particles
  which are not necessarily in $\L_n$ (unless $t=0$).  \*

Call $\r_n(t),V_n(t),\NN_n(t)$ the quantities in Eq.\equ{e5.2}
evaluated for
$\lis x^{(n,0)}(t)$ and let $\lis \r_n(t),$ $\lis V_n(t),\lis\NN_n(t)$ be
the corresponding quantities defined by taking the maxima and minima
in the full $\L_n$ ({\it i.e.} not excluding points in $\O_0$).

For times $0\le t\le \Th$ consider {\it $\L_n$-regularized}
motions (see Sec.\ref{sec3}) evolving from $x$ with $n$ fixed (see
comment (6), p.\pageref{t}). Define

\be R_n(t)\defi n^\b +\int_0^t V_n(s) \frac{v_1\,ds}{r_\f},
\Eq{e5.4}\ee
where $R_n(0)=g_{1/d}(2^n)=n^{1/d}$ for $d=1,2$ but $R_n(0)=n^{1/2}$
for $d=3$, and $v_1\,V_n(s)$ is the maximum
speed that a moving particle {\it inside a thermostat} can acquire in the
time interval $[0,s]$ under the $\L_n$-regularized evolution: formally
$v_1V_n(s)=\max_{q_{i}\not\in\O_0,s'\le s} |\dot q_{i}(s') |$.

By the choice of $\b$ it is $R_n(0)\ge n^{1/{d}}$ for
$d=1,2,3$: hence it will be possible to claim if $d=1,2$ that
$W(x(0);\x,R_n(0))\le n\,\EE(x(0))<\infty$ with $\m_0$--probability
$1$, see Eq.\equ{e3.10},\equ{e3.11} and Appendix B. If $d=3$ the
somewhat weaker bound $W(x(0);\x,R_n(0))\le n^{3/2}\,\EE(x(0))<\infty$
will hold.

The dimensionless quantity $R_n(t)$ will also provide a convenient
upper bound to the maximal distance a moving particle inside any
thermostat can travel during time $t$, in units of $r_\f$, following
the $\L_n$--regularized motion.

Then the following {\it a priori energy bound}, proved in Appendix C,
holds: \*

\0{\bf Theorem 2:} {\it The $\L_n$--regularized frictionless dynamics
satisfies, for $t\le\Th$ and if $W(x,R)\defi\sup_\x W(x;\x,R)$:

\be W(\lis x^{(n,0)}(t), R_n(t))\le \,C\, R_n(t)^d \Eq{e5.5}\ee
for $n$ large enough and $C>0$ (depending only on $\EE$).
} \*

\noindent{\it Remarks:} (1) hence we obtain also a bound on the force
that can be exercised on a particle by the others {\it including the
forces due to the particles in $\O_0$} (as the latter force is
bounded proportionally to $N_0$, hence by a constant):

\be \sum_j |F_{ij}|\le CR_n(t)^{d/2} \Eq{e5.6}\ee
\\
(2) the inequality holds $\forall d$'s. The bounds in
Eq.\equ{e5.5}, \equ{e5.6} will be repeatedly used.
\*

By the definition Eq.\equ{e3.10} of $W$ it follows that

\be V_n(s)\le \,C\, R_n(s)^{d/2}\Eq{e5.7}\ee
{\it Therefore} for $d=1,2$, going back to Eq.\equ{e5.4} and solving
it, $R_n(t)$ is bounded proportionally to $R_n(0)=n^{1/d}$: hence
$R_n(t)\le C n^{1/d}$ for $0\le t\le \Th,\,d=1,2$ and $W\le C n$ hence
$ V_n(t)\le C n^{1/2}$ and the bounds in the first line of
Eq.\equ{e5.8}, in theorem 3 below, are immediate consequences.

\*\0{\bf Theorem 3:} {\it It is $\r_n(t)\ge \lis \r_n(t),V_n(t)\le
\lis V_n(t),\NN_n(t)\le \lis\NN_n(t)$ and up to time $t\le \Th$ the
following inequalities hold for $d=1,2$ or $d=3$ respectively:

\be \kern-1mm\eqalign{ \lis V_n(t)\le& C\, n^{\frac{1}2},\quad\lis
\NN_n(t)\le C n^{\frac12},\quad \lis \r_n(t)\ge \, c\,
n^{-\frac1\a}\cr \lis V_n(t)\le& C\, n^{\frac{1}2},\quad\lis
\NN_n(t)\le C n^{\frac34},\quad \lis \r_n(t)\ge \, c\,
n^{-\frac3{2\a}}\cr} \Eq{e5.8}\ee }
\*

The $d=3$ case is more delicate because the inequality obtained from
Eq.\equ{e5.4} using Eq.\equ{e5.7} gives a blow-up in a finite
time. 

Following \cite{CMP000} we shall prove in Appendices D,E
(theorem 7) that Eq.\equ{e5.7} can be improved to $V_n(t)\le C
R_n(t)$, hence $V_n(t),R_n(t)\le C n^{1/2}$, which implies the second
of Eq.\equ{e5.8} for $V_n(t),\NN_n(t),\r_n(t)$. The distinction
between $\lis \r_n(t),\lis V_n(t), \lis\NN_n(t)$ and
$V_n(t),\NN_n(t),\r_n(t)$ is only really necessary in the $d=3$ case.

For $d=1,2$, the bounds on $V_n(t),\NN_n(t),\r_n(t)$ in the first line
Eq.\equ{e5.8} have been just discussed as a corollary of theorem 2.
For $d=3$ the bounds in Eq.\equ{e5.8} will be proved for
$V_n(t),\NN_n(t),\r_n(t)$ first and then they will be shown to imply
the bounds in Eq.\equ{e5.8} for the $\lis \r_n(t),\lis V_n(t),
\lis\NN_n(t)$: see Appendix E, paragraphs after Eq.\equ{e9.26}

%%%%%%%%%%%%%%%%%%%%%%%%%%%%%%%%%%%%%%%%%%
%%%%%%%%%%%%%%%%%%%%%%%%%%%%%%%%%%%%%%%%%%
\def\SEC{\small\SECsei}
\section{\SECsei}\label{sec6}
%Infinite volume. Frictionless dynamics
\iniz
%%%%%%%%%%%%%%%%%%%%%%%%%%%%%%%%%%%%%%%%%%
%%%%%%%%%%%%%%%%%%%%%%%%%%%%%%%%%%%%%%%%%%

It will now be checked that the $n\to\infty$ limit motion
exists in the sense of the local dynamics assumption, {\it i.e.}
existence of $\lis x^{(0)}(t)\defi\lim_{n\to\infty}$ $
\lis x^{(n,0)}(t)$ and a suitable form of its uniqueness.

The equation of motion, for
a particle in the $j$-th container (say), can be written as

\be \lis q_i^{(n,0)}(t)=q_i(0)+t \dot q_i(0) +
\int_0^t(t-\t)\,f_i(\lis x^{(n,0)}(\t))\,d\t\Eq{e6.1}\ee
\noindent{}where the label $j$ on the coordinates (indicating the
container) is omitted and $f_i$ is the force acting on the selected
particle divided by its mass (for $j=0$ it includes the stirring
force).

Existence of the dynamics in the frictionless, open boundary case
will be discussed proving

\* \0{\bf Theorem 4:} {\it If $d\le 3$ and $x\in \HH_{1/d}$ the
thermodynamic limit evolution $\lis x^{(0)}(t)_i=\lim_{n\to\infty}
\lis x^{(n,0)}(t)_i$ exists.}
\*

\0{\it Proof:} (adapted from the proof of theorem 2.1 in
\cite[p.32]{CMP000}, which applies essentially
unaltered). Let

\ifthenelse{\formato=4}{
\be\eqalign{ \d_i(t,n)\defi&|\lis q_i^{(n,0)}(t)-\lis q^{(n+1,0)}_i(t)|,\cr
u_k(t,n)\defi&\max_{q_i\in \L_k}\d_i(t,n),\cr}\Eq{e6.2}\ee}
{\be\d_i(t,n)\defi|\lis q_i^{(n,0)}(t)-\lis q^{(n+1,0)}_i(t)|,\qquad
u_k(t,n)\defi\max_{q_i\in \L_k}\d_i(t,n),\Eq{e6.2}\ee}
then Eq.\equ{e6.1} yields

\ifthenelse{\formato=4}{
\be \eqalign{\d_i(t,n)\le&  \int_0^t \frac{\Th}{m}\,d\t\ \big\{
F'_w\d_i(\t,n)\cr
&+F'\sum_j
(\d_j(\t,n)+\d_i(\t,n))\,\big\}\cr}\Eq{e6.3}\ee}
{\be \d_i(t,n)\le \int_0^t \frac{\Th}{m}\,d\t\ \big\{
F'_w\d_i(\t,n)+F'\sum_j
(\d_j(\t,n)+\d_i(\t,n))\,\big\}\Eq{e6.3}\ee}
where $F'_w= C \frac{\f_0}{ r^2_\ps}\,
n^{\frac{\a+2}\a}$ bounds the maximum gradient of the walls plus the
stirring forces (see Eq.\equ{e5.8}) for $d=1,2$ and for $d=3$
we can take $F'_w= C \frac{\f_0}{ r^2_\ps}\,
n^{\frac32\frac{\a+2}\a}$; $F'=\max_q|\partial^2 \f(q)|$; and
the sum is over the number $\lis \NN_n$ of the particles $\lis
q_j(\t)$ that can interact with $\lis q_i(\t)$ at time $\t$. The
latter, by Eq.\equ{e5.8}, is $\lis\NN_n\le Cn^{1/2}$
($d=1,2$) or $\le C n^{3/4}$ ($d=3$) for both $\lis x^{(n,0)}(\t)$ and
$\lis x^{(n+1,0)}(\t)$. Let

\be \h\defi
(\frac32+\frac3\a),\;\; 2^{k_1}\defi 2^{k}+r_n
\Eq{e6.4}\ee
where $r_n r_\f$ is the maximum distance a particle can travel in time
 $\le\Th$, bounded by Eq.\equ{e5.8} by $C\, r_\f\,n^{1/2}$ (for $d=2$
 the $\h$ could be taken $\h=(1+\frac2\a$)). Then

\be \frac{u_k(t,n)}{r_\f}\le \,C\, n^\h \int_0^t
\frac{u_{k_1}(s,n)}{r_\f} \frac{ds}\Th\Eq{e6.5}\ee
($C$ is a function of $\EE$ as agreed in Sec.\ref{sec5}).
Eq.\equ{e6.5} can be iterated $\ell $ times; choosing $\ell$ so that
$2^k+ C\, \ell n^{1/2} <2^{n}$, {\it i.e.}
$\ell=\frac{2^{n}-2^k}{2C n^{1/2}}$ which is $\ell\,> \,c \, 2^{n/2}\,
\d_{k<n}$ for $n$ large.

By
Eq.\equ{e5.8} $u_n(t,n)$ is $\le C\, n^{1/2}$, so that for  $n>k$,

\be\frac{ u_k(n,t)}{r_\f}\le
\, C'\, \frac{(n^{\h})^{\ell+1}}{\ell!} n^{1/2}
\le C\, 2^{- 2^{n/2}c } \Eq{e6.6}
\ee
for suitable $C',C,c>0$ ($n$--independent functions of $\EE$). Hence the
evolutions locally ({\it i.e.} inside the ball $\L_k$) become closer
and closer as the regularization is removed ({\it i.e.}  as
$n\to\infty$) and very fast so.

If $q_i(0)\in\L_k$, for $n>k$ it is

\be \lis q_i^{(0)}(t)=\lis q_i^{(k,0)}(t)+\sum_{n=k}^\infty
(\lis q^{(n+1,0)}_i(t)-\lis q^{(n,0)}_i(t))
\Eq{e6.7}\ee
showing the existence of the dynamics in the thermodynamic limit
 because also the inequality,  for $n>k$,

\be\frac{|\dot {\lis q}^{(n,0)}_i(t)-\dot
{\lis  q}^{(n+1,0)}_i(t)|}{v_1}\le \,C\, 2^{- 2^{n/2}c }\Eq{e6.8}\ee
follows from Eq.\equ{e6.6} and from $ \dot {\lis q}^{(n,0)}_i(t)-\dot
{\lis q}^{(n+1,0)}_i(t)=\int_0^t\big( f_i(\lis q^{(n,0)}(\t))
-f_i(\lis q^{(n+1,0)}(\t))\big)d\t$. Or, for $n>k$ and $|q_i(0)|<r_\f
2^k$,

\be|\lis x^{(n,0)}(t)_i-\lis x^{(n+1,0)}(t)_i|\le \,C\, e^{-c
2^{n/2}}\Eq{e6.9}\ee
calling $|x_i-x_i'|\defi {|\dot q_i-\dot q'_i|}/{v_1}+
{|q_i-q'_i|}/{r_\f}$.

Hence existence of the thermodynamic limit dynamics in the
frictionless case with open boundary conditions is complete and it
yields concrete bounds as well.
Uniqueness follows from Eq.\equ{e6.3}: we skip details, \cite{CMS005}.
Hence we have obtained:\*

\noindent{\bf Theorem 5:} {\it There are $C(\EE),c(\EE)^{-1}$,
increasing functions of $\EE$, such that the frictionless evolution
satisfies the local dynamics property and if $q_i(0)\in\L_k$

\be \eqalign{
&|\dot {\lis q}^{(n,0)}(t)| \le v_1 C(\EE) \, k^{\frac12} ,\cr
&{\rm distance}( \lis q_i^{(n,0)}(t), \partial(\cup_j \O_j\cap \L))\ge
c(\EE) k^{-\frac{3}{2\a}} r_\ps\cr
&\NN_i(t,n)\le C(\EE) \, k^{3/4}\cr
&|\lis x^{(n,0)}_i(t) -\lis x^{(0)}_i(t)|
\le C(\EE) r_\f e^{-c(\EE) 2^{nd/2}}\cr}
\Eq{e6.10}\ee
$\forall\,n> k$. The $\lis x^{(0)}(t)$ is the unique solution of the
frictionless equations satisfying the first three of Eq.\equ{e6.10}.}
\*

For $d=1,2$ in the above proof the bounds in the first line of
    Eq.\equ{e5.8} could be used instead of the weaker ones in the
    second line (which holds also for $d=3$): in this way the
    exponents $\frac{3}{2\a}$ and $\frac34$ can be improved to
    $\frac1{\a}$ and $\frac12$ respectively.

It would also be possible to show the stronger result that $\lis
x^{(0)}(t)\in \HH_{1/d}$ for $d=1,2$ and $\lis x^{(0)}(t)\in
\HH_{3/2}$ for $d=3$: but, for the proof of theorem 1, theorems 2-5
are sufficient, hence the proof of the stronger property is relegated
to theorem 11 in Appendix F.

The corresponding proof for the thermostatted evolution will be
somewhat more delicate: and {\it it will be weaker} as it will not hold
under the only assumption that $\EE(x)<\infty$ but it will be
necessary to restrict further the initial data to a subset of the
phase space (which however will still have $\m_0$--probability $1$).

%%%%%%%%%%%%%%%%%%%%%%%%%%%%%%%%%%%%%%%%%%
%%%%%%%%%%%%%%%%%%%%%%%%%%%%%%%%%%%%%%%%%%
\def\SEC{\small \SECsette}
\section{\SECsette}\label{sec7}
%Entropy bound. Thermostatted dynamics
\iniz
%%%%%%%%%%%%%%%%%%%%%%%%%%%%%%%%%%%%%%%%%%
%%%%%%%%%%%%%%%%%%%%%%%%%%%%%%%%%%%%%%%%%%

In this section the regularized motion, thermostatted and with {\it
elastic boundary conditions}, will be compared to the thermodynamic
limit of the frictionless dynamics with {\it open boundary conditions}
whose existence and main properties have been established in
Sec.\ref{sec6}.

A direct comparison of the two evolutions along the lines of
  Sec.\ref{sec6} does not seem possible because the quantity $\a_j$ is
  not small enough compared to the ``Lyapunov exponent'' $n^\h$ in
  Eq.\equ{e6.5}: the extra $\a_j$ introduces a small inhomogeneous
  term in attempting a derivation of the analogue of
  Eq.\equ{e6.5}. The extra term is of order $2^{-nd}$ which is
  amplified at rate $Cn^\h$, $\h>1$ by Eq.\equ{e6.4}, to a quantity
  $O(2^{-nd} e^{C\Th n^\h})$ diverging with $n$.

The strategy will be to introduce a stopping time $T_{n}(x)$ in the
$\L_n$--regularized dynamics defined by the time in which either the
total kinetic energy in any thermostat or a local energy reaches a
conveniently fixed threshold. The threshold will be defined so that
within $T_{n}(x)$ the $\a_j$ are  small and the frictionless and
thermostatted dynamics are very close at least well inside $\L_n$. We
shall then conclude  that the number of particles $M$ and their maximal speed
$Vv_1$ in a
region $\DD$ of width $r_\f$ adjacent to $ \O_0$ are, in the
thermostatted dynamics, bounded in terms of the
respective values  in the frictionless motion with the same
initial conditions (finite as shown in Theorem 3 for $\EE(x)\le
E$). See Eq.\equ{e7.3}, \equ{e7.4} below.

Thus {\it before the
stopping time} $T_n(x)$ the entropy production in the
thermostatted motion is simply controlled (by Eq.\equ{e3.4} it is $\le
C\, M\, V $), showing that the measure $\mu_0$ is ``quasi invariant''.
This will imply that with large $\mu_0$ probability the thermostatted motion
cannot
have a too short stopping time and, actually, that the stopping time
cannot be reached before $\Th$. The precise statements can be found in
the proof of theorem 6.

Restricting attention to the set $\XX_{E}\subset \HH_{1/d}$ of initial
data $\XX_{E}\defi\big\{x\,|\, \EE(x)\le E;\, \big\}$ the constants
$C,C',\dots,$ $c,c',\ldots$ will be functions of $E$ as stated in
Sec.\ref{sec5}.

Let $\r_{\O_0}(\x)$ be the  distance  of $\x\not\in\O_0$
from the boundary $\partial\O_0$ and let

\be\kern-2.mm{  \L_*\defi\kern-.5mm \{\x: \r_{\O_0}(\x)\le r_\f\},\,
\L_{**}\defi\kern-.5mm \{\x: \r_{\O_0}(\x)\le 2r_\f\}\
}\Eq{e7.1} \ee
By the result in Sec.\ref{sec6} there are $M$ and $V$ (which depend on
$E$) so that for all $x\in \XX_E$, with the notations
Eq.\equ{e3.9}, for $n$ large enough:

\ifthenelse{\formato=4}
{\be\eqalign{ &\max_{t\le \Th} N_{\L_{**}}(\lis S^{(n,0)}_{t}x)
<M \cr &\max_{t\le \Th} \,V_{\L_{**}}(\lis S^{(n,0)}_{t}x)< V-1
\cr}\Eq{e7.2}\ee}
{\be\max_{t\le \Th} N_{\L_{**}}(\lis S^{(n,0)}_{t}x)
<M,\qquad \max_{t\le \Th} \,V_{\L_{**}}(\lis S^{(n,0)}_{t}x)< V-1
\Eq{e7.2}\ee}

Fix, once and for all, $\k>0$ smaller than the minimum of the kinetic
energy densities of the initial $x$ in the $\n$ thermostats
($x$-independent with $\m_0$--probability $1$, by the ``no phase
transition'' assumption, {\it i.e.} by ergodicity, see comment (c)
p.\pageref{npt}).

Let $C_\x$ the cube with side $r_\f$ centered at a point $\x$ in
the lattice $r_\f \ZZZ^d$, and using the definitions in
Eq.\equ{e3.9}

\be\|x\|_n\defi\max_{\x\in\L_n} \frac{ \max
(N_{C_\x}(x),{\e_{C_\x}(x)})}{g_{1/2}(\x/r_\f)},\Eq{e7.3}\ee
where $\e_{C_\x}(x)\defi\sqrt{e_{C_\x}(x)}$. Fixed $\g$ once and for
all, arbitrarily with $\frac12<\g<1$, define the stopping times

$$\eqalignno{T_{n}&(x)\defi \max\big\{t : \,t\le
  \Th:\, \forall\, \tau< t,&\eq{e7.4} \cr &
  \frac{K_{j,n}(S^{(n,1)}_\t x)}{\f_0}\,>\k
  2^{nd},\quad
    \|S^{(n,1)}_tx\|_{n} < (\log n)^{\g}\big\}. \cr}$$

\noindent{\bf Theorem 6:} {\it If $d\le 3$, $\exists\,\, C,C',c>0$
 depending only on $E$ such that for all $n$ large enough:
\\
(1) For all $x\in \XX_E$, $t\le T_n(x)$ and all
 $q_i(0)\in \L_{(\log n)^\g}$

\be \eqalign{ &|q^{(n,a)}_i(t) -\lis q^{(n,0)}_i(t)|\le \,C\, r_\f\,
e^{-(\log n)^\g\,c },\cr &|\dot q^{(n,a)}_i(t) -\dot {\lis
q}^{(n,0)}_i(t)|\le \,C\,v_1\, e^{-(\log n)^\g\,c }.  \cr}
\Eq{e7.5}\ee
furthermore with the notations
Eq.\equ{e3.9}, for $n$ large enough and for all ${t\le T_n(x)}$:

\be N_{\L_{*}}(S^{(n,1)}_{t}x)
\le M,\ V_{\L_{*}}(S^{(n,1)}_{t}x)\le  V
\Eq{e7.6}\ee
(2) the $\m_0$-probability of the set $\BB\defi\{x\,|\, x\in \XX_E$
    {\rm and} $T_n(x)\le\Th\}$ is

\be \m_0(\BB) \le \,C\,e^{-c (\log n)^{2\g}+C' M V}.\Eq{e7.7}\ee
}

{\it Assuming theorem 6} it is immediate to see that it implies the
  assumptions of theorem 1 (locality of the dynamics) because $\g>1/2$
  in Eq.\equ{e7.7} will imply (by Borel--Cantelli's lemma) that, with
  $\m_0$--probability $1$, eventually for large $n$, $T_n(x)>\Th$ and
  therefore, by Eq.\equ{e7.5}, the thermodynamic limits of
  $x^{(n,a)}(t),\lis x^{(n,0)}(t)$ will coincide for $t\le\Th$. Hence
  {\it theorem 6 implies theorem 1}.  \*

\noindent{\it Remarks:} (1) The proof of theorem 6 contains a standard
part, namely proving Eq.\equ{e7.5} and its corollary Eq.\equ{e7.6}
{\it for $t\le T_n(x)$}: it follows from the existing literature, see
\cite{FD977,CMP000} and the proof in Section \ref{sec6}, and for completeness
it is discussed, providing
the details, in appendix K.  \\
(2) The proof of Eq.\equ{e7.7}, {\it i.e.} the proof that the
Eq.\equ{e7.6} is not an empty statement, {\it is the technically
  original contribution of this paper}. It is based on an ``entropy
estimate'' and the strategy of its use is discussed below, deferring
to the Appendix J the actual computations to obtain the estimate.
\\ (3) The proof below refers only to the case $a=1$: the same
argument applies for the (easier case of) frictionless motions with
elastic boundary conditions, $a=0$.  \*

\0{\it Proof:} The stopping time $T_n(x)$ is the time when the
trajectory $S_t^{(n,1)}x$, $x\in \XX_E$, crosses the (piecewise
smooth) surface $\Si'$ of points $y$ where either the kinetic energy
in some $\Omega_j\cap \Lambda_n$ has the value $\kappa 2^{nd}$ or 
$\|S_{t}^{(n,1)}y\|_n$ crosses from below the value $(\log n)^\g$.
%Denote by $\Si$ the intersection with $\Si'$ of all  trajectories
%$S_t^{(n,1)}x$, $x\in \XX_E$,  with $T_{n}(x)\le \Theta$.  
Denote by $\Si$ the subset of $\Si'$ of all points
$S_{T_{n}(x)}^{(n,1)}x$, $x\in \XX_E$,   $T_{n}(x)\le \Theta$. Thus
$\Si= \cup_{\t\le\Th}S_{\t}^{(n,1)}\Si_\t$ where
\be\Si_\t\defi\big\{ x\in\XX_E\,| \ T_{n}(x)=\t\big\}
\Eq{e7.8}\ee
and $\BB$ is the disjoint union of $\Si_\t$
over $\tau \le \Theta$.

The surfaces $\Sigma$, $\Si'$ and $\Si_\t$, $\t<\Th$, are
symbolically described in Fig.2.

\eqfig{180}{86}{%
\ins{61}{75}{$x$}
\ins{-1}{57}{$\Si_\t$}
\ins{6}{39}{$\Si_{\t'}$}
\ins{61}{20}{$S^{(n,1)}_{-\th(x)}x$}
\ins{120}{72}{$\Si'\supset \Si$}
}{fig2}{}

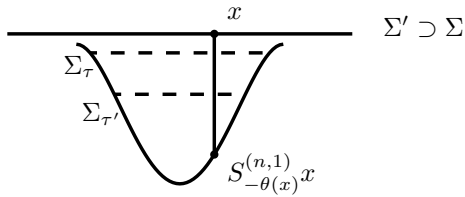
\noindent{Fig.2: \small\it The horizontal ``line'' represents $\Si'$,
$\Si$ is the subset of $\Si'$ under the ``curve''
$S^{(n,1)}_{-\th(x)}x$; $\theta(x)\le\Th$ is such that
$S^{(n,1)}_{-\th(x)}x\in \XX_E$ while $S_{-t}^{(n,1)}x\notin \XX_E$
for all $t\in (\theta(x),\Theta]$; the vertical line represents the
trajectory of the point $S^{(n,1)}_{-\th(x)}x$, the incomplete
(``dashed'') lines the ``levels'' $\Si_\t, \Si_{\t'}$, their missing
parts are made of points not in ${\cal X}_E$ but with an ``ancestor''
in $\mathcal X_E$.  } \*

This is the setup described in Appendix H, with $\Si$ a ``base'' and
$\th(x)=\max_{\t\le \Th} $ $\big\{\t\,|\,
S_{-\t}^{(n,1)}x\in\Si_\t\big\}$ a ``ceiling function'', so that (see
Appendix H and, for early applications of this key estimate, see
\cite{Si974,MPP975, FD977}):

 \be
 \mu_0( \mathcal B )\le\int_{\Si}
 \n_{0,\Si}(dy) \int_{0}^{\theta(y)}dt\,\,w(y)\,\,
e^{\widehat\sigma(y,t)}
\Eq{e7.9}\ee
where (1) $\n_{0,\Si'}$ denotes the surface area measure (measuring
the volume with $\m_0$) on $\Si'$,
\\
(2) $w=|v_x\cdot n_x|$ with $v_x$ the $\dot x^{(n,1)}$ evaluated at $x$
(by the equations of motion)  and $n_x$ is the normal to $\Si$ at $x$,
\\
(3) $\widehat\sigma(y,t)\defi$ $
\int_{-t}^0|\sigma(y^{(n,1)}(t'))|\,dt'$ bounds the phase space
contraction, {\it i.e.} the entropy production.
\*

Effectively this means that the distribution $\m_0$ can be treated as
an invariant one for the purpose of estimating probabilities in $\BB$
via Eq.\equ{e7.9}, because phase space contraction $\s=\s(x)$ is given
by Eq.\equ{e3.3} and $Q_0$, see comment (2) following Eq.\equ{e3.3},
can be estimated in the same way as $Q_j$, by $C'M V$ ({\i.e} by a
bound on the speed times the number of particles in $\L_*$ times the
number of particles in $\O_0$). It follows that the integral
$\int_0^{T_{n}(x)}|\s(x^{(n,a)}(t))|\,dt$ is also uniformly bounded
(in $\BB$) by $C MV$. Therefore a volume element in $\BB$ contracts at
most by $e^{- C M V}$ on the trajectory of $\m_0$-almost all points
$x\in\BB$, up to the stopping time $T_{n}(x)$, and it follows

\be
 \mu_0( \mathcal B )\le e^{C M V}\int_{\Si'}
|w(y)| \n_{0,\Si'}(dy) \Eq{e7.10}
\ee
which we call, naturally, {\it entropy bound}.

The derivation of the Eq.\equ{e7.7} becomes, by Eq.\equ{e7.10}, an
``equilibrium estimate'', as it does not involve times $t>0$ and is a
standard consequence of the superstability property of the potential
$\f$: the detailed calculations are in Appendix J.

\* \0{\it Remark:} The analysis in this section is valid for all
dimensions except for the use of the properties of $\lis x^{(0)}(t)$,
theorem 6, which have been derived only for $d\le 3$.

%%%%%%%%%%%%%%%%%%%%%%%%%%%%%%%%%%%%%%%%%%
%%%%%%%%%%%%%%%%%%%%%%%%%%%%%%%%%%%%%%%%%%
\def\SEC{\small\SECotto}
\section{\SECotto}\label{sec8}
%Conclusions
\iniz
%%%%%%%%%%%%%%%%%%%%%%%%%%%%%%%%%%%%%%%%%%
%%%%%%%%%%%%%%%%%%%%%%%%%%%%%%%%%%%%%%%%%%

Equivalence between different thermostats is widely studied in the
literature and the basic ideas, extended here, were laid down in
\cite{ES993}. A clear understanding of the problem was already set up
in comparing isokinetic, isoenergetic and Nos\'e-Hoover bulk
thermostats in \cite{ES993}, where a history of the earlier results is
presented as well, see also \cite{Ru000, Ru00b,Ga008d}.

There are, since a long time, studies of systems with free
thermostats, starting with \cite{FV963}. Such thermostats are somewhat
pathological and may not always lead to the stationary states that
would be expected: as exemplified in the case of simple spin chains,
\cite{ABGM972,Le971}.  More recently similar or identical thermostat
models built with free systems have been considered starting with
\cite{EPR999}.

Isokinetic thermostats should be treated in a very similar way,
\cite{Ga008d}: the extra difficulty is that the entropy production in
a finite time interval receives a contribution also from the time
derivative of the total energy of the reservoirs, \cite{Ga008d}, and
further work seems needed.

More general cases, like Lennard-Jones interparticle potentials are
difficult, see \hbox{\cite{BGGZ005}}, and new ideas may be needed.

Finally here the interaction potential has been assumed smooth:
singularities like hard core could be also considered at a heuristic
level. It seems that in presence of hard cores plus smooth repulsive
potentials all estimates of Sec.\ref{sec5} are still valid
but the existence of the limiting motion as $\L\to\infty$ remains a
difficult point because of the discontinuities in the velocities due
to collisions.

It should be noted that the key bounds Eq.\equ{e5.8} hold for the open
boundary conditions motions $\lis x^{(n,0)}(t)$ (as needed for our
purposes) and have {\it not} been proved for $x^{(n,a)}(t)$. A careful
analysis of our argument shows that the bounds hold only for particles
initially in $\L_k$, $k<n$, in the cases of reflecting boundary
conditions, but not for $k=n$: the results of Sec.\ref{sec7} show that
the maximal speed $V_n(t)$ is bounded by $ C\,n^{1/2}(\log n)^\g$ in
the $\L_n$-regularized thermostatted or frictionless dynamics rather
than by $C n^{1/2}$, which is the bound obtained for the frictionless
dynamics with open boundary conditions.

The main problem left open is what can be said about the limit
$t\to\infty$, {\it i.e} the study of the stationary state reached at
infinite time. A conjecture has been proposed, \cite{Ru002}, that can
be interpreted as saying that if $d=1,2$ the limit will be
``trivial'': {\it i.e} it will be an {\it equilibrium Gibbs
distribution} at some intermediate temperature if $\BF=0$. But, again
interpreting the conjecture in \cite{Ru002}, for $d=3$ the stationary
distribution(s) will be nontrivial and asymptotically they will be
Gibbs distributions at the initial temperatures and densities of the
thermostats.

\ifthenelse{\formato=5}{\*\0{\bf Acknowledgements:} This work has
been partially suppor\-ted
also by Rutgers University.}{}
\ifthenelse{\formato=4}
{}{\vfill\eject}
%%%%%%%%%%%%%%%%%%%%%%%%%%%%%%%%%%%%%%%%%%

\def\inizA{\setcounter{equation}{0}{
\rhead{\thepage}\lhead{{{{\small\bf\thesection:}\ \SEC}}}}
\renewcommand{\theequation}{\Alph{section}.\arabic{equation}}
}

%%%%%%%%%%%%%%%%%%%%%%%%%%%%%%%%%%%%%%%%%%
%\def\SEC{Appendices}
\appendix
%\section{Appendices}\label{sec9}\inizA
%%%%%%%%%%%%%%%%%%%%%%%%%%%%%%%%%%%%%%%%%%
%%%%%%%%%%%%%%%%%%%%%%%%%%%%%%%%%%%%%%%%%%
\newcounter{appendicep}%\setcounter{paragrafo}{27}
\def\piuap{{\addtocounter{appendicep}{1}}}%\Alph{appendicep}.}
%%%%%%%%%%%%%
%\centerline{\small\bf{\piuap \ \  Appendix: \AppA}}
%\*\*
\def\SEC{\small Appendix: \AppA\piuap }\inizA
\section{Superstability. Sets of full measures}\label{A}
%%%%%%%%%%%%%

In our simple case ($\f\ge0$, decreasing and $r_\f<\infty$)
the superstability bound is reduced to Schwartz' inequality.
In fact, by the dimensionless energy definition in Eq.\equ{e3.10},
$W\ge (\frac {U}{\f_0}+N)\ge \frac{\lis \f}{2\f_0} \sum _p N_p^2$ with
the sum running over labels $p$ of disjoint boxes of diameter
$\frac{r_\f}2$ covering a cube or a ball $\D$ and containing $N_p\ge0$
particles (in particular: $N=\sum_p N_p$), hence over $\ell
\le{|\D|}{(2\sqrt{d}/r_\f)^{d}}$ terms.  By the Schwartz' inequality
$\sqrt{{\lis \f}/{2\f_0}} N\le \sqrt{W\,\ell}$, hence Eq.\equ{e5.1}.\*

There are $c_0$ and $R_0$ and a strictly positive, non decreasing
function $\g(c)$, $c\ge c_0$, so that $\forall c\ge c_0,\forall R\ge
R_0$,

\be \mu_0\Big( \{W(x,0,R) \ge c R^d\}\Big) \le e^{-\g(c) R^d} \Eq{e9.1}
\ee
(which more generally holds for superstable but not necesssarily
positive $\f$, \cite{Ru970}).

If $g:\ZZZ^d \to \RRR_+$, $g(i)\ge 1$, $c\ge c_0$, the probability

\be \mu_0\Big( \cap _{i\in \ZZZ^d,r\in\ZZZ, r \ge g(i)}\{W(x; i, r) \le
  c\,r^d\} \Big) \Eq{e9.2} \ee
is $\ge 1- \sum_{i\in \ZZZ^d, r\in\ZZZ,r \ge g(i)} e^{-\g(c) r^d}$
with the sum being bounded proportionally to the sum $ \sum_{i\in
\ZZZ^d} e^{-\g(c) [g(i)]^d} $ which converges if $g(i) \ge
c'(\log_+|i|)^{1/d}$, with $c'$ large enough.  \*\*

%%%%%%%%%%%%%%%%%%%%%%%%%%%%%%%%%%%%%%%%%%%%%%%%%%%%%%%%%%%%%%%
\def\SEC{\small Appendix: \AppB}\label{B}\inizA
%\centerline{\small\bf{\piuap \ \  Appendix: \AppB }}\*\*
\section{Choice of $R_n(t)$\piuap}%\label{B}
%%%%%%%%%%%%%%%%%%%%%%%%%%%%%%%%%%%%%%%%%%%%%%%%%%%%%%%%%%%%%%%

The proof of the inequalities Eq.\equ{e5.5} yields $\forall t\le
\Theta$ that $ W(\lis S^{(n,0)}_t x,R) \le c W(x ,R +\int_0^t V_n(\t)d\t/\Th)
$ provided $R$ is such that $ \frac{R +\int_0^t V_nd\t/\Th}{R} \le 2$,
which is implied by $ R\ge R_0 + \int_0^t V_n(s)ds/\Th,\ R_0 \ge 0 $.
The maximal speed $v_1V_n(t)$ at time $t$ is bounded by $V_n(t) \le \sqrt
{W(\lis S^{(n,0)}_t x,R)}$.  The choice $R_0=R_n(0)=n^{1/d}$  is the weakest
that still insures that the set of initial data has $W(x;0,R)/R^d$
finite with $\m_0$--probability $1$, see Appendix A.
\*\*

%%%%%%%%%%%%%%%%%%%%%%%%%%%%%%%%%%%%%%%%%%
%%%%%%%%%%%%%%%%%%%%%%%%%%%%%%%%%%%%%%%%%%
\def\SEC{\small Appendix: \AppC}\label{C}\inizA\piuap
%\centerline{\small\bf{\piuap \ \  Appendix: \AppC}}
\section{Theorems 2 ($d\le3$) and 3 ($d\le2$)}
%%%%%%%%%%%%%%%%%%%%%%%%%%%%%%%%%%%%%%%%%%
%%%%%%%%%%%%%%%%%%%%%%%%%%%%%%%%%%%%%%%%%%
\*\*

Consider the ball $\BB(\x, R_{n}(t,s))$
around $\x\in\RRR^d$ of radius

\be R_n(t,s)\defi R_n(t)+\int_s^t \frac{v_1\,V_n(s)}{r_\f}\,ds\ge
1.\Eq{e9.3}\ee
The ball radius shrinks as $s$ increases between $0$ and
$t$ at speed $v_1\,V_n(s)$: therefore no particle can enter it.

The quantity $R_n(t,s)$ can be used to obtain a bound on the size of
$W(\lis x^{(n,0)}(t);\x,R_n(t))$ in terms of the initial data
$x(0)=x$ and of

\be W(x,R)\defi\sup_{\x} W(x;\x,R).\Eq{e9.4}\ee
Remark that $W(\lis x^{(n,0)}(\t),R)$ is finite if the just defined
meaning of $\lis x^{(n,0)}(\t)$ is kept in mind (so that $\lis
x^{(n,0)}(\t)$ contains finitely many particles).

Let $\ch_\x(q, R)$ be a smooth function of $q-\x$ that has value $1$
in the ball $\BB(\x,R)$ and decreases radially to reach $0$ outside
the ball $\BB(\x,2R)$ with gradient bounded by $2(r_\f\,R)^{-1}$. Let
also

\ifthenelse{\formato=4}
{\be\eqalign{\widetilde W(x;\x,R)&\defi\frac1{\f_0}
\sum_{q\not \in\O_0}\ch_\x(q,R)\cr &\kern-9mm
\cdot \big(\frac{m\dot q^2}2+\ps(q)+
\frac12\sum_{q'\not \in\O_0,\,q'\ne q}\f(q-q')+\f_0\big),\cr
&\widetilde W(x;R)\defi\sup_{\x} \widetilde W(x;\x,R).\cr
}\Eq{e9.5}\ee}
{\be\eqalign{&\widetilde W(x;\x,R)\defi\frac1{\f_0}
\sum_{q\not \in\O_0}\ch_\x(q,R)\cdot \big(\frac{m\dot q^2}2+\ps(q)+
\frac12\sum_{q'\not \in\O_0,\,q'\ne q}\f(q-q')+\f_0\big),\cr
&\widetilde W(x;R)\defi\sup_{\x} \widetilde W(x;\x,R).\cr}\Eq{e9.5}\ee}
Denoting $B$ an estimate of how many balls of radius $1$ are needed in
$\RRR^d$ to cover a ball of radius $3$ (a multiple of the radius large
enough for later use in Eq.\equ{e9.9}) so that every pair of points at
distance $<1$ is inside at least one of the covering balls, it follows
that $W(x;\x,2R)\le B\, W(x,R)$, see Eq.\equ{e9.4}, so that:

\ifthenelse{\formato=4}
{\be\eqalign{& W(\lis x^{(n,0)}(\t);\x,R)\le\widetilde W(\lis
x^{(n,0)}(\t);\x,R)\cr
&\kern1cm\le W(\lis x^{(n,0)}(\t);\x,2R),\cr & \widetilde
W(x;\x,R)\le \,B\, W(x,R)\cr}\Eq{e9.6}\ee}
{\be \eqalign{& W(\lis x^{(n,0)}(\t);\x,R)\le\widetilde W(\lis
x^{(n,0)}(\t);\x,R) \le W(\lis x^{(n,0)}(\t);\x,2R),\cr
&\widetilde
W(x;\x,R)\le \,B\, W(x,R)\cr}\Eq{e9.6}\ee}
Although $W$ has a direct physical interpretation $\widetilde W$
turns out to be mathematically more convenient and
equivalent, for our purposes, because of  Eq.\equ{e9.6}.

The proof of theorem 2 is taken, from the version in {\rm
\cite[p.34]{CMP000}} of an idea in {\rm \cite[p.72]{FD977}}: it is
repeated for completeness because quite a few minor modifications are
needed here. It could be extended, essentially unaltered to the other
boundary conditions, see \cite[Sec.6-A]{GP009}, but we shall not need
it here.

Consider $\widetilde W(\lis x_{\L_n}(s);\x,R_n(t,s))$, for $0\le s\le
t\le \Th$:

\be\eqalign{ &\frac{d}{ds} {\widetilde W}(\lis x_{\L_n}(s);\x,R_n(t,s))\le
\sum_{q\not \in\O_0} \frac{\ch_\x(q(s),R_n(t,s))}{\f_0}\cr
&\cdot\frac{d}{ds} \big(\frac{m\dot q(s)^2}2+\ps(q(s))+
\frac12\sum_{q'\not \in\O_0}\f(q(s)-q'(s))\big)\cr}\Eq{e9.7}\ee
because the $s$--derivative of $\ch_\x(q(s),R_n(t,s))$ is $\le0$ since
no particle can enter the shrinking ball $\BB(\x,R_n(t,s))$ as $s$
grows: {\it i.e.} $\ch_\x(q(s),R_n(t,s))$ cannot increase with $s$.
The sums are restricted to the $q$'s of $\lis x_{\L_n}(s)$.

In the frictionless case a computation of the derivative in
Eq.\equ{e9.7} leads, with the help of the equations of motion and
setting $\ch_\x(q(s),R_n(t,s))\equiv\ch_{\x,q,t,s}$, to

\be\eqalign{ & \frac{d}{ds} \widetilde W(\lis x_{\L_n}(s);\x,R_n(t,s))\le
\sum_{q\not\in\O_0} \frac{\dot q(s) F(x(s))}{\f_0}\ch_{\x,q,t,s}\cr &
-\sum_{q,q'\not \in \O_0}\big(\ch_{\x,q,t,s}\,-
\ch_{\x,q',t,s}\,\big) \frac{\dot q(s)\partial_q
\f(q(s)-q'(s))}{2\f_0}\cr }\Eq{e9.8}\ee
where $F(x(s))$ denotes the force that the particles in $\O_0$
exercise on the thermostats particles and the dot indicates a
$s$--derivative; keep in mind that positions and
velocities of particles outside $\L_n$ are not considered, in the
$\L_n$--regularized dynamics at open boundary conditions.

Since the non zero terms have $|q(s)-q'(s)|<r_\f$, the gradient
of $\ch$ is $\le2\, (r_\f R_n(t,s))^{-1}$ and $|\dot q|,|\dot q'|\le v_1
V_n(s)=r_\f |\dot R_n(t,s)|$ it follows, since $F=\max(N_0|\partial
\f|)$ is an upper bound on the force that particles in $\O_0$
can exercise on a particle in the thermostats,

$$\eqalignno{ &\frac{d}{ds} \widetilde W(\lis
x_{\L_n}(s);\x,R_n(t,s))\le \frac{F v_1}{\f_0} \widetilde W(\lis
x_{\L_n}(s);\x,R_n(t,s)) \cr & +\frac{F r_\f}{\f_0} \frac {|\dot
R_n(t,s)|}{R_n(t,s)}B\, \widetilde W(\lis x_{\L_n}(s);\x,2
R_n(t,s)+1)&\eq{e9.9}\cr &\le B^2\, \frac{F v_1}{\f_0} (\frac{r_\f}{
v_1}\frac {|\dot R_n(t,s)|}{R_n(t,s)}+1)\widetilde W(\lis
x_{\L_n}(s);R_n(t,s))\cr}$$
where $\widetilde W(x;R)$ is defined in analogy with Eq.\equ{e9.4},
and the first term in the {\it r.h.s.} is obtained by bounding the
corresponding term in Eq.\equ{e9.8} as $\sum_q (|\dot q|\sqrt\ch)
\sqrt\ch$ followed by Schwartz' inequality.

By Eq.\equ{e9.9},\equ{e9.6}, $R_n(t,s)/R(t,0)\le 2$, $\wt W(\lis
x_{\L_n}(s),$$\x,$ $ R_n(t,s))$ $\le\wt W(\lis x_{\L_n}(s),R_n(t,s))$
we get the inequality

\be W(\lis x_{\L_n}(s),R_n(t,s))\le \,C\,
W(\lis x_{\L_n}(0), R_n(t,0))\Eq{e9.10}\ee
with $C\defi \exp\big({(\frac{Fr_\f}{\f_0}\log2+\frac{3F
v_1}{\f_0})\,\Th\, B}\big)$.

This concludes the proof of theorem 2.
\*

To prove theorem 3 for $d=2$ the inequalities Eq.\equ{e5.8} for
$\r_n(t),V_n(t),\NN_n(t)$ follow from $R_n(t)\le C n^{1/d}$ and from
Eq.\equ{e5.5},\equ{e5.7}. It remains to check them for $\lis
\r_n(t),\lis V_n(t),\lis\NN_n(t)$, {\it i.e.} it re\-mains to consider
the particles moving inside $\O_0$.

The inequalities certainly extend to the initial time $t=0$ and to any
$t$ for $\lis\NN_n(t)$ because the number of particles in $\O_0$ is
fixed and bounded in terms of $\EE$. The variation $E(t)-E(0)$ of the energy
contained inside the region $\O_0$ equals the work $\LL_{ext}$ done by the
particles outside $\O_0$ plus the work $\LL_\F$ of the stirring forces
in the time $t$. By the bounds Eq.\equ{e5.6},\equ{e5.7}
this implies, for $t\le \Th$:

$$ \eqalignno{ &\sum_{q\in\O_0} \big(\frac m2 \dot q(t)^2+\ps
(q(t))\big)\le E(0)+ (\frac {N_0^2}2+N_0 n^{1/2}) C
\cr&
+\int_0^t(N_0\|\F\| \max_{q\in\O_0} |\dot q(\t)|+N_0 n^{1/2}\, C\,
V_n(\t))d\t &\eq{e9.11}\cr}$$
Therefore if $v(t)=\max_{q\in\O_0, t'\le t}|\dot q(t')|$ it follows that $
v(t)^2\le C' n+ C'' v(t)$ which implies $v(t)\le v_1 C n^{1/2}$; hence
$\ps(q(t)\le C n$ and the Eq.\equ{e5.8} are checked for $d=2$.

Theorem 3 will be complete after the proof of the second line of
Eq.\equ{e5.8} which follows from the speed bounds in Appendix E.  \*

\*\*
%%%%%%%%%%%%%%%%%%%%%%%%%%%%%%%%%%%%%%%%%%%%%%%%%%%%%%%%%%%%%%%
\def\SEC{Appendix: \AppD}\label{D}\inizA
%\centerline{\small\bf\piuap\ \ \AppD}
\section{\piuap Work bounds. Thermostats in $d=3$}
%%%%%%%%%%%%%%%%%%%%%%%%%%%%%%%%%%%%%%%%%%%%%%%%%%%%%%%%%%%%%%%
\*\*

{\it In this and and in the next appendix we study the $3$-d dynamics
$\lis S^{(n,0)}_tx\equiv\lis x^{(n,0)}(t)\equiv (\lis
q_h^{(n,0)}(t),\dot{\lis q}_h^{(n,0)}(t))$, frictionless and with open
boundary conditions.}  We follow the method of \cite{CMP000}.  \*

Time intervals will be often subdivided into subintervals of equal
length $\d$: in general, however, the number may be not an integer. We
shall intend, without mention, that the last of the intervals might be
shorter. This is used, without mention, only in cases in which the
bounds considered remain valid also for the ``left over'' interval.

Let $\e(\dot q,q)\defi \sqrt{\frac1{\f_0}(\frac{m}2\dot q^2+\ps(q))}$,
abridged as $\e_j(t)\defi$ $
\e(\dot {\lis q}_j(t),{\lis q}_j(t))$ when $q,\dot q$ are time
dependent. Remark that the speed of a particle is $\le v_1\e(q,\dot
q)$ and its energy in the field of the walls forces is
$\frac{mv_1^2}2\e(\dot q,q)^2$.

A particle $q_i\not\in\O_0$ will be called {\it fast} in a time
interval $J\subset [0,t]$ if $\e_{min}\defi \min_{\t\in
J}|\e_j(\t)|\ge R_n(t)$.

The main property on which the analysis rests is an estimate by $C
R_n(t)^{2-\g}$ of the work ${\cal L}\defi\int_J\sum_j \dot q_j F_{ij}$
that the $i$-th particle, assumed ``fast'', performs over the
neighbors in the ``short'' time interval $J=[s-\th,s]$, with $\th\defi
\frac{r_\f}{v_1} R_n(t)^{-\g}$ with $s\le t$ and $1>\g>0$,
\cite{CMP000}.   And the precise property that we
shall need is expressed by the lemma:

\* \0{\bf Lemma 1:} {\it Consider the dynamics $\lis S^{(n,0)}_t$ in
the geometry of Fig.1 and in a time interval $J\subset[0,\Th]$ of size
$\th =\frac{r_\f}{v_1} R_n(t)^{-\g}$. Then the work ${\cal
L}\defi\int_J\sum_j \dot q_j F_{ij}$ that the $i$-th particle, assumed
``fast'' and in $\O_j,\, j>0$, performs over the neighbors in the time
interval $J$ is bounded by $|\LL|\le C R_n(t)^{2-\g}$ if
$\g\in(\frac12,\frac23]$ and $J\subset[0,t]$.}  \*

This will be proved considering {\it the special geometry in Fig.1},
following \cite{CMP000}.

 \* {\it All things considered it will turn
out that if $d=3$ the parameter $\g$ can be arbitrarily fixed in the
interval $(\frac12,\frac23]$.}  \*

\0{\it Remark:} The restriction on the geometry is severe: it is used
only to prove lemma 5 below. It is very likely, however, that lemma 5
holds for {\it all thermostats considered in Sec.\ref{sec1}, and {\rm
that} our equivalence holds in the general cases as well.}
\*

Notice that all particles $q_j$ interacting with a given particle at
a given time $\t$ are contained inside a ball of radius $2r_\f
R_n(\t)$.

Identifying particles with their labels and ima\-gi\-ning the
following maxima to be taken over ${\t\in J}$, divide the particles
$(q_j,\dot q_j)$ into three groups, \cite{CMP000}, as ``slow'',
``intermediate'', ``quick'': \*

\halign{#&\ #\hfill\cr
(a)& ${\cal A}_0=\, \hbox{\it slow} = \{j| \max|\e_j(\t)|\le
  R_n(t)^\g\}$,\cr
\noalign{\*}
(b)& ${\cal A}_k=$ {\it intermediate} =
$ \{j| 2^{k-1}R_n(t)^\g< \max|\e_j(\t)|\le$\cr
& $2^{k} R_n(t)^\g\},$ for $k=1,\ldots,k_m-1$, with $k_m$ being\cr
&determined by the condition $2^{k_m} R_n(t)^{\g}= R_n(t)$,\cr
\noalign{\*}
(c)& ${\lis A}=\, \hbox{\it quick} = \{j|
\max|\e_j(\t)|>\frac12 R_n(t)\}$,\cr}
\*

Correspondingly the work ${\cal L}={\cal L}_0+\sum_k{\cal L}_k+\lis{{\cal L}}$:
the three contributions will be estimated separately.

This corresponds to proposition 4.3 in \cite{CMP000}.  The
time interval $J=[s-\th,s]$ ($s\ge\th$) is assumed, for
simplicity, to be of length $\th=R_n(t)^{-\g}r_\f/v_1$.
The estimates, however, hold also for shorter intervals.  \*

\*\*
%%%%%%%%%%%%%%%%%%%%%%%%%%%%%%%%%%%%%%%%%%
\centerline{\small\bf \Alph{appendicep}-1
Work $\LL_0$ of fast particles on slow particles}
\label{subsec6-A}
%%%%%%%%%%%%%%%%%%%%%%%%%%%%%%%%%%%%%%%%%%
\*\*

Divide the time interval $J$ into $H$ consecutive intervals
  $\D_h=[s_{h},s_{h+1}]$ of size equal to $\frac{r_\f}{2 v_1R_n(t)}$ and let
  $H=2R_n^{1-\g}$ be their number.  The work $\LL_0$ is bounded by

\ifthenelse{\formato=4}{\be \eqalign{
|\LL_0|\le& \|\partial\f\| v_1 R_n(t)^\g\int_J N(\t)d\t\cr
\le&C  R_n(t)^\g\sum_{h=1}^H |\D_h| \max_{\t\in\D_h}
N(\t)\cr}\Eq{e9.12}\ee}
{\be
|\LL_0|\le \|\partial\f\| v_1 R_n(t)^\g\int_J N(\t)d\t
\le C  R_n(t)^\g\sum_{h=1}^H |\D_h| \max_{\t\in\D_h}
N(\t)\Eq{e9.12}\ee}
where: $N(\t)$ is the number of slow particles in the ball of radius
$r_\f$ and center $\lis q_i^{(n,0)}(\t)$.  The time intervals of size
$\th=R_n(t)^{-\g}r_\f/v_1$ are so short that a slow particle cannot
travel a distance greater than the range $r_\f$ away from
$q_j(s-\th)$.

A slow particle contributes to $N(\t)$ if $|\lis q_j^{(n,0)}(\t)-\lis
q_i^{(n,0)}(\t)|\le r_\f$, hence $|\lis q_j^{(n,0)}(s-\th)-\lis
q_i^{(n,0)}(\t)|\le 2 r_\f$ because $|\lis q_j^{(n,0)}(s-\th)-\lis
q_j^{(n,0)}(\t)|\le r_\f$.

If $N_h\defi \max_{\t\in\D_h} N(\t)$ then:

\be |\LL_0|\le C R_n(t)^{-(1-\g)/2} (\sum_k N_h^2)^{1/2}\Eq{e9.13}\ee
having used $\sum_{h=1}^H N_h\le \sqrt{H}\sqrt{\sum_h N_h^2}$ and
$H=2R_n^{1-\g}$.

Notice that $N_h^2$ is bounded by $W(\lis x(s-\th)), $ $\lis
q_i^{(n,0)}(\t),2)$ where $\t$ is the time when the maximum defining
$N_h$ is reached.

Hence, if $\TT_h$ denotes the tube spanned by the ball of radius
$2r_\f$ and center at $\lis q_i^{(n,0)}(t)$ for $t\in \D_h$, $N_h^2$
is bounded by $W(\TT_h)$ provided $W(\G)$ is defined as in
Eq.\equ{e3.10} with $\BB(\x,R)$ replaced by $\G$.

\*

\0{\bf Lemma 2:} {\it Each $\TT_h$ can be intersected by at most $C$
other $\TT_{h'}$.}
\*

\0{\it Remark:} Thus $\TT_h$ is a set contained in a ball of radius
$r_\f R_n(t)$ (which is, by definition, the maximum distance {\it any}
particle can travel) and each point is in at most $C$ different
$\TT_h$'s. This implies $\sum_h W(\TT_h)\le C\,W(\TT)\le C\,\EE\,
R_n(t)^d$, if $
 \TT\defi\cup_h\TT_h$. Concluding: $(\sum_h N_h^2)^{1/2}\le C\, W(\TT)^{1/2}$ so
that by Eq.\equ{e9.13} it is $|{\cal L}_0|\le C\f_0
R_n^{-\frac{1-\g}2}R_n^{{d}/2}$, $d=3$, because $\TT$ is contained in
the ball of radius $R_n(t)\,r_\f$ centered at $\lis
q_i^{(n,0)}(s-\th)$, or

\be |{\cal L}_0|\le
C\f_0 R_n(t)^2 R_n^{-(1-\frac\g2)}\le C\f_0 R_n(t)^{2-\g}
\Eq{e9.14}\ee
by Eq.\equ{e5.5}, provided
$1-\frac\g2\ge\g$. This is possible for $d=3$ if $ \g\le
\frac23$.  \*

\0{\it Proof:} To check lemma 2 let $\t\to\l_0(\t)$ be the
path that the fast particle would follow under its own inertia and the
walls forces, which act in a small strip $\DD$ of width $r_\ps<r_\f$
near the containers walls, in a time interval of size $\th=R_n(t)^{-\g}
r_\f/v_1$.

Given a point $\x\in\O_j$ the particle following $\l_0(\t)$ might
spend up to two intervals of time $J_1,J_2\subset J$ inside the ball of
radius $r_\f$ centered on $\x$ (possibly one before the collision with
the wall and one after bouncing). However the
actual path $\t\to\l(\t)$ differs from $\l_0$.

Suppose that initially the fast particle is outside $\DD$: until
outside $\DD$ it proceeds, in time $\le\th$, over a distance

\ifthenelse{\formato=4}{\be\eqalign{
\ell(\t)\ge& |\dot q_i(s_{h})|(\t-s_{h})-\int_{s_{h}}^\t (\t-s)
\frac{|F_i(s)|}m\,ds\cr
\ge& |\t-s_h|\,(v_{\min}-C v_1 R_n(t)^{-\g+d/2}),\cr}\Eq{e9.15}\ee}
{\be
\ell(\t)\ge |\dot q_i(s_{h})|(\t-s_{h})-\int_{s_{h}}^\t (\t-s)
\frac{|F_i(s)|}m\,ds
\ge |\t-s_h|\,(v_{\min}-C v_1 R_n(t)^{-\g+d/2}),\Eq{e9.15}\ee}
by Eq.\equ{e5.6}, undergoing a deflection by an angle which is at most
$C R_n(t)^{-\g+d/2}/R_n(t)$ and will have velocity $\ge \frac{v_1}2
R_n(t)$, for $n$ large.

At the entrance into the region $\DD$ at a time $t_1$ the difference
between $\l(t_1)$ and $\l_0(t_1)$ will be $\le |t_1-s_h|\,C\, v_1
R_n(t)^{-\g+d/2}$ and there are two possibilities for $n$ large. The
component of the velocity along the normal to the wall is $\ge
\frac1{\sqrt2}v_1 \frac{R_n(t)}2$ or its tangential component is $\ge
\frac1{\sqrt2}v_1 \frac{R_n(t)}2$.

In the first case the time spent by $\l_0$ inside $\DD$ will be $\le
\frac{4\sqrt2 r_\ps}{v_1} R_n(t)^{-1}$ and then at a time $t_2$ the
particle will be again out of $\DD$ with a position
$|\l(t_2)-\l_0(t_2)|\le (t_2-s_h) \,C\, v_1\, R_n(t)^{-\g+d/2}$ and
speed normal to the wall $> c\, R_n(t)$. The motion will then proceed
a distance bounded as in Eq.\equ{e9.15} (with a larger $C$).

In the second case the tangential velocity is large and the particle
will move away from the entrance point into $\DD$ keeping its velocity
component parallel to the wall within $\le C
R_n(t)^{-\g+d/2}/R_n(t)$ dashing away from any fixed ball of radius
$r_\f$, without  coming close to any point within $2r_\f$ more than
twice.

Next we can consider the case of a fast particle initially in $\DD$:
if the velocity component normal to the wall is $\ge \frac1{2\sqrt2} v_1
R_n(t)$ the particle will get out of $\DD$ in a short time $<C
\frac{r_\ps}{v_1}R_n(t)^{-1}$; if the tangential component is large
the particle will move away from initial position in $\DD$, keeping
its velocity component parallel to the wall and undergoing a deflection
$\le C R_n(t)^{-\g+d/2}/R_n(t)$.
\*

\0{\it Remark:} {\it Lemma 2 is likely to hold under the sole
assumption that the fast particle path $\l_0$ moves away from the
origin at radial speed bounded below proportionally to the initial
value of $\e_i$. Hence it holds probably in full generality for the
thermostats considered in Sec.\ref{sec1}}.

\*\*
%%%%%%%%%%%%%%%%%%%%%%%%%%%%%%%%%%%%%%%%%%
\centerline{\bf \small\bf \Alph{appendicep}-2
Work $\LL_k$ of fast particles on}
\centerline{\small\bf intermediate speed particles
}\label{subsec6-B}
%%%%%%%%%%%%%%%%%%%%%%%%%%%%%%%%%%%%%%%%%%

\*\* If $N(\t)$ is the number of intermediate speed particles within
an interaction radius of $\lis q_i(\t)$, ${\cal L}_k$
is bounded by

\ifthenelse{\formato=4}{\kern-3mm }{}
\be |\LL_k|\le ||\partial\f|| v_1 2^k R_n(t)^\g \int_{s-\th}^s
N(\t)\,d\t\Eq{e9.16}\ee
analogously to Eq.\equ{e9.12}.

The relative speed of the particles $q_i$ and $q_j$, by the definition
of $\AA_k$, can be bounded below by $|\e_i(\t)-\e_j(\t)|\ge R_n(t)-
2^{k_m-1} R_n(t)^\g\ge\frac12 R_n(t)$ for $n$ large as long as the
fast particle is outside $\DD$.

The contribution of the $j$-th particle to ${\cal L}_k$ lasts,
therefore, a time bounded by $C R_n(t)^{-1}$ except for the time, also
of the order of $R_n(t)^{-1}$, spent by the fast particle within the
wall with tangential speed smaller than a fraction of $R_n(t)$, say
$3R_n(t)/4$. If at $\t_j\in J$ the $\e_j(\t)$ is maximum, then

\ifthenelse{\formato=4}{\be\eqalign{ &|{\cal L}_k|\le |{\cal A}_k|\,
C\, R_n(t)^{-1}\, 2^{k}\, R_n(t)^{\g},\qquad{\rm and}\cr &|{\cal
A}_k|2^{2(k-1)} R_n^{2\g}\le \sum_{j\in{\cal A}_k} \e_j(\t_j)^2\cr
&\le \sum_{j\in{\cal A}_k}\Big( \e_j(s-\th)^2 + C \int_{s-\th}^s d\t
|\dot q_j(\t)| N_j(\t)\Big) \cr}\Eq{e9.17}\ee}
{\be\eqalign{ &|{\cal L}_k|\le |{\cal A}_k|\,
C\, R_n(t)^{-1}\, 2^{k}\, R_n(t)^{\g},\qquad{\rm and}\cr &|{\cal
A}_k|2^{2(k-1)} R_n^{2\g}\le \sum_{j\in{\cal A}_k} \e_j(\t_j)^2\le
\sum_{j\in{\cal A}_k}\Big( \e_j(s-\th)^2 + C \int_{s-\th}^s d\t
|\dot q_j(\t)| N_j(\t)\Big) \cr}\Eq{e9.17}\ee}
where $N_j(\t)$ is the number of particles of $\lis x^{(n,0)}(\t)$ which interact
with $q_j(\t)$. We multiply both both sides of \equ{e9.17} by
$2^{-k}$, bound $|\dot q_j(\t)| 2^{-k}\le v_1 R_n(t)^\g$ and

\be \sum_k\sum_{j\in \AA_k} N_j(\t)\le C R_n(t)^d,\qquad
d=3\Eq{e9.18}\ee
which can be proved as follows: let $\BB',\BB$ be the balls with some
center $q_i(s-\th)$ and radius $2 R_n(t)$ and $3R_n(t)$,
respectively. Then

\ifthenelse{\formato=4}{\be
\eqalign{
&\sum_k \sum_{j\in \AA_k} N_j(\t)\le \sum_{j:\,q_j(\t)\in\BB'}
\sum _{
\ell\,: | q_\ell(\t)-q_j(\t)|\le r_\f} 1\cr
&\le \sum^* N_{C_i}N_{C_j}\cr}\Eq{e9.19}\ee}
{\be\sum_k \sum_{j\in \AA_k} N_j(\t)\le \sum_{j:\,q_j(\t)\in\BB'}
\sum _{
\ell\,: | q_\ell(\t)-q_j(\t)|\le r_\f} 1\le \sum^* N_{C_i}N_{C_j}
\Eq{e9.19}\ee}
where the $C_i$'s are elements of a covering of $\BB$ by disjoint cubes
of diameter $r_\f/2$ and $N_{C_i}$ are the numbers of particles in the
$C_i$'s; and the $\sum^*$ denotes sum over $C_i,C_j\subset\BB$ with $dist
(C_i,C_j)\le r_\f$. The inequality continues as

\ifthenelse{\formato=4}{\be
\eqalign{
&\sum^*\frac12(N_{C_i}^2+N_{C_j}^2)\le C \sum_{C_i\subset \BB} N_{C_i}^2\cr
&\le W(x(\t),q_i(s-\th), R_n(t))\le C R_n(t)^d
\cr}\Eq{e9.20}\ee}
{\be
\sum^*\frac12(N_{C_i}^2+N_{C_j}^2)\le C \sum_{C_i\subset \BB} N_{C_i}^2
\le W(x(\t),q_i(s-\th), R_n(t))\le C R_n(t)^3
\Eq{e9.20}\ee}
(by Eq.\equ{e5.5}) which proves Eq.\equ{e9.18}. Since $\sum_k$
$2^{-k} $
$\sum_j\e_j(\t)^2\le C R_n(t)^3$, by %S 
Eq.\equ{e5.5}, we get from Eq. \equ{e9.17}), using Eq.\equ{e9.18}:

\ifthenelse{\formato=4}{
\be\eqalign{ &\sum_k|{\cal A}_k| 2^{k-2} R_n^{2\g}\le \sum_k 2^{-k}
\sum_{j\in {\cal A}_k}\e_j(s-\th)^2 \cr & + C' R_n(t)^\g R_n(t)^3
R_n(t)^{-\g}\le C^{''} R_n(t)^3. \cr}\Eq{e9.21}\ee}
{\be\sum_k|{\cal A}_k| 2^{k-2} R_n^{2\g}\le \sum_k 2^{-k}
\sum_{j\in {\cal A}_k}\e_j(s-\th)^2 + C' R_n(t)^\g R_n(t)^3
R_n(t)^{-\g}\le C^{''} R_n(t)^3. \Eq{e9.21}\ee}
Therefore $\sum_k |{\cal A}_k| 2^k\le \lis C (R_n^{3-2\g})\le C
R_n(t)^{3-\g}\th/\Th$ and $\sum _k {\cal L}_k$ is bounded, via Eq\equ{e9.17},
proportionally  to

\be R_n(t)^\g R_n(t)^{-1} R_n(t)^{3-\g} \frac{\th}\Th= C
R_n(t)^{2-\g}\Eq{e9.22}\ee

\*\*
%%%%%%%%%%%%%%%%%%%%%%%%%%%%%%%%%%%%%%%%%%
\centerline{\small\bf \Alph{appendicep}-3
Work $\lis {{\cal L}}$ of a fast particle on quick particles}
%\centerline{\small\bf  and speed estimate}\label{subsec6-C}
%%%%%%%%%%%%%%%%%%%%%%%%%%%%%%%%%%%%%%%%%%

\*\* The work (of the inner forces) on the quick particles can be
bounded by first remarking that their $\e_j(\t)^2$ at a time $\t\in J$
is bounded below by $ \frac14R_n(t)^2 $ minus $C
R_n(t)^{\frac{d}2-\g}\e_j$ (upper bound on the work done by the other
particles in time $\frac{r_\f}{v_1}R_n(t)^{-\g}$). Therefore it
follows that $\e_j(\t)\ge c\, R_n(t)$ for $n$ large enough, since
$\g>\frac12$.

So that $|\lis {{\cal A}}| \big(c \,R_n(t)\big)^2\le
\sum_j|\dot q_j(\t)|^2\le C v_1^2 R_n(t)^3$, by Eq\equ{e5.5}.

It follows that $|\lis {{\cal A}}|\le C \EE R_n(t)$. Therefore
 by the
energy estimate (see Eq.\equ{e5.5}) an application of Schwartz'
inequality bounds $|\lis\LL|$ above by

\be \le C' |\lis {{\cal A}}|^{\frac12}\int_{s-\th}^sd\t ( \sum_{j\in
\lis {{\cal A}}} \e_j(\t)^2)^{1/2} \le C
R_n(t)^{\frac12}R_n(t)^{\frac32}\th\Eq{e9.23}\ee
by Eq.\equ{e5.5}, {\it i.e.}  $|\lis{{\cal L}}| \le C  R_n(t)^{2-\g}$.

\*\*
%%%%%%%%%%%%%%%%%%%%%%%%%%%%%%%%%%%%%%%%%%%%%%%%%%%%%%%%%%%%%%%
%%%%%%%%%%%%%%%%%%%%%%%%%%%%%%%%%%%%%%%%%%%%%%%%%%%%%%%%%%%%%%%
\def\SEC{Appendix: \AppE}\label{E}\inizA\piuap
%\centerline{\small\bf\piuap\ \ Appendix: \AppE}\*\*
\section{Speed bounds in $d=3$}
%%%%%%%%%%%%%%%%%%%%%%%%%%%%%%%%%%%%%%%%%%%%%%%%%%%%%%%%%%%%%%%
%%%%%%%%%%%%%%%%%%%%%%%%%%%%%%%%%%%%%%%%%%%%%%%%%%%%%%%%%%%%%%%
\*\*

We can now infer an estimate for maximal speed, $v_1V_n(t)$, and
 maximal distance, $R_n(t) r_\f$, that a thermostat particle
can travel in the $\L_n$-regularized motion.  \*

\0{\bf Theorem 7:} {\it Given $x\in\HH_{1/d}$, $d=3$, there is a
constant $G$ depending only on $\EE=\EE(x)$ and $\Th$
such that for any particle of $x$ initially in $\L_n$ it is
$\e(t)< %\sqrt{\frac{m v_1^2}2}
G
\,R_n(t)$ for all $t\le \Th$.
Therefore, by Eq.\equ{e5.4}, $V_n(t)\le C R_n(t)$ and
$R_n(t)\le C R_n(0)\equiv C\,n^{1/2}$ for $t\le \Th$.}  \*

\0{\it Proof.} Suppose that there are $t^*\le t\le \Th$ and
$q_i(0)\in\L_n$
so that  $\e(t^*)=G R_n(t)$. We claim that the above cannot hold if
$G^2>\frac43 C''$, where $C''$ is a constant defined below, in the course of
the proof.

Remark that the claim implies theorem 7 with $G=(4 C'')/3)^{1/2}$.
And the claim is proved as follows. Let $C_1$ be such that $\e_i(0)\le
C_1 R_n(0)$ for all $x$ with $\EE(x)\le E$. Then by choosing $C''$
large enough it will be $G\ge 2 C_1$.  Since $\e_i(t^*)= G R_n(t)> C_1
R_n(t)\ge \e_i(0)$ there exists $t_1<t^*$ and $\e_i(\t)\ge C_1 R_n(t)$
for all $\t\in [t_1,t^*]$. We shall next prove that $t^*-t_1\ge C
R_n^{-1/2}$ so that $[t_1,t^*] $ splits into the union of $H$
intervals of length $\th=\frac{r_\f}{v_1} R_n(t)^{-\g}$ with $H$ large. In fact
$\frac{d \e_i(\t)^2}{d\t}$ $\le C \e_i(\t)\frac1{2m} |F_i(x(\t))|$, hence
$\frac{d \e_i(\t)}{d\t}\le C\,\frac1{2m}| F_i(x(\t))|$, and by
Eq.\equ{e5.6} (for non integer $H$ see the comment at the beginning of
Appendix D):

\be G R_n(t)\le C_1 R_n(t)+C (t^*-t_1)
R_n(t)^{{d}/2}\Eq{e9.24}\ee
which yields $t^*-t_1\ge c
R_n^{-\frac12}\gg R_n^{-\g}\Th$ if $\g>\frac12, \,d=3$.

The variation $(\e_i(t^*)^2-\e_i(t_1)^2)$ equals the work of the pair
interaction forces over the $i$-th particle. It is therefore the
variation of the pair-potentials energy of the particle $q_i$ in the
configurations $\lis x^{(n.0)}(t_1),\lis x^{(n.0)}(t^*)$ {\it not
taking into account} the $N_0$ particles in $\O_0$, plus the work that
the particles in $\O_0$ perform on the particle $q_i$, plus the work
that particle $i$ performs on the neighbors. The three terms are bounded
by

\*
\0(a) $||\partial\f||$ times the number of particles
interacting with $i$ at the two times, hence by $C R_n(t)^{{d}/2}$ and
by Eq.\equ{e5.6},\equ{e5.5} (valid also if $d=3$, see Sec.\ref{sec5}),
\*

\0(b) the work performed on the particle $q_i$ by the $N_0$ particles in
$\O_0$ is bounded by $C N_0 R_n(t)^{d/2}$ by Eq.\equ{e5.5},
\*
\0(c) $|\sum_{h=1}^H \int_{t_1+(h-1)\th}^{t_1+h\,\th}
\sum_{j}\dot q_j(\t) F_{ij}(x(\t))d\t|$
\*

\0so that, by the work estimates
and by $H\th\le \Th$, it follows

\be
G^2 R_n(t)^2\le C_1^2 R_n(t)^2 +  C'' \,R_n(t)^2
\Eq{e9.25}\ee
for $n$ (hence $R_n(t)$) large enough if $C''$ is the largest of
all constants met so far. Since $G\ge 2C_1$, Eq.\equ{e9.25} implies
$G^2\le \frac43 C''$ and the proof of theorem 7 is complete.

Going back to the definition of $R_n(t)$ in Eq.\equ{e5.4} it follows
that $\exists \,C>0$, depending only on $\EE$, such that for $t\le\Th$,

\ifthenelse{\formato=4}{
\be \eqalign{ &W(\lis x^{(n,0)}(t),R_n(t))\le C R_n(t)^d,\qquad \cr
&R_n(t)\le \,C\, n^{{1}/2},\qquad V_n(t)\le C
n^{1/2},\cr }\Eq{e9.26} \ee}
{
\be W(\lis x^{(n,0)}(t),R_n(t))\le C R_n(t)^d,\qquad
R_n(t)\le \,C\, n^{{1}/2},\qquad V_n(t)\le C
n^{1/2},\Eq{e9.26} \ee}
which means that the maximum speed and the maximum distance a particle
{\it not in $\O_0$} can travel in the $\L_n$-regularized dynamics
grows as a power of $n$, at most.

It remains to consider the particles in $\O_0$ to derive the bounds
Eq.\equ{e5.8} with $V_n(t),$ $\NN_n(t),\r_n(t)$ replaced by $\lis
V_n(t),\lis\NN_n(t),\lis\r_n(t)$, {\it i.e} by the quantities in which
the particles in $\O_0$ are also taken into account. However the proof
of the corresponding statement for $d=1,2$ (see paragraphs preceding
and following Eq.\equ{e9.11} in Appendix C) applies unaltered as it
only depends on bounds on $V_n(t)$ proportional to $n^{1/2}$.

It can be remarked that the Eq.\equ{e9.26} summarizes results that in
the cases $d=1,2$ were the contents of theorems 3,4 in
\cite{GP009}. However the bound on $R_n(t)$ is substantially worse
than the corresponding in dimensions $1,2$ because it only allows us to
bound the maximum number of particles that can fall inside a ball of
radius $r_\f$ by $C n^{d/4}$ rather than by $C n^{1/2}$ as in Eq.\equ{e5.8}.

A key implication of the above results is a proof of Theorem 4 in
$d=3$:

\*\0{\bf Theorem 8:} {\it If $x\in \HH_{1/d}$ the
thermodynamic limit evolution $\lis x^{(0)}(t)_i=\lim_{n\to\infty}
(\lis S^{(n,0)}_t x)_i$ exists.}
\*

\0{\it Proof:} Remark that in Sec.\ref{sec6} no use has been made
of entropy bounds. The only difference is that with the new bound in
Eq.\equ{e9.26} $\h$ in Eq.\equ{e6.4} is one
unit larger.

\*
By the final remark to Sec.\ref{sec7} the analysis
leads to
\*\*

\0{\bf Theorem 9:} {\it With $\m_0$--probability $1$ in $x\in
\HH_{1/d}$, $d=3$: the frictionless motions have the local dynamics
property and, as $n\to\infty$, the limits of $\lis S^{(n,0)}_t x,
S^{(n,1)}_t x$ exist and coincide, and furthermore the Eq.\equ{e2.3}
with $\L=\RRR^d$ for $\lis x^{(0)}(t)$ have a unique solution for data
in $\HH_{1/3}$ and with values in $\HH_{3/2}$.}  \*

This is proved but for the uniqueness statement: as in
$d=1,2$ the proof is by iteration of the integral
equations for two solutions with the initial data in $\HH_{\z}$ and
with values in $\HH_{9\z/2}$: we skip the details, see \cite{CMS005}.

The analysis in Sec.\ref{sec7} can also be followed word by word to
check the expected result \*

\0{\bf Theorem 10:} {\it The limits $\lim_{n\to\infty} S^{(n,0)}_t x$ exist
and coincide with $\lim_{n\to\infty}
\lis S^{(n,0)}_t x$.}
\*

The check is left to the reader.
\*\*
%%%%%%%%%%%%%%%%%%%%%%%%%%%%%%%%%%%%%%%%%%%%%%%%%%%%%%%%%%%%%%%
\def\SEC{\small Appendix: \AppF}\label{F}\inizA
%\centerline{\small\bf\piuap\ \  Appendix: \AppF}
\section{Frictionless motion is a flow if $d=2$}\piuap
\*\*
%%%%%%%%%%%%%%%%%%%%%%%%%%%%%%%%%%%%%%%%%%%%%%%%%%%%%%%%%%%%%%%

The following theorem is obtained by a straightforward adaptation of
theorem 2.2 in \cite{CMS005}.  \*

\noindent{\bf Theorem 11:}  {\it Let $E>0$ and $d=1,2$. Then,
given any $\Th$ there is $E'$ (depending on $\Th,E$)
so that for all $x$ such that $\EE(x)\le E$

\be \EE(\lis S^{(0)}_tx )\le E',\qquad \hbox{for all}\
  t\le\Th\Eq{e9.27}\ee
so that the evolution $x\to \lis S_t^{(0)}x$ is a flow in
$\HH_{1/d}$. For $d=3$ data in $\HH_{1/3}$ evolve into data in
$\HH_{3/2}$.  Hence the same holds for $S^{(a)}_tx$ for $a=0,1$ by
theorem 7.}  \*

\noindent{\bf Proof.} The proof is based on the bounds
Eq.\equ{e6.8},\equ{e6.9} derived in the proofs of the theorems 6,7
(for $d=1,2$) and 10 (for $d=3$).

Let $\lis x^{(n,0)}_t\defi \lis S^{(n,0)}_t x$
and consider

\be \widetilde W\big(\lis x^{(0)}_t, \x,\r \big),\quad {\rm for}\quad
\r \ge (\log_+ ({|\x|}/{r_\f}))^{1/d} \Eq{e9.28} \ee
with $\wt W$ defined as in Eq.\equ{e9.5}.  Let $n_\x-1$ be the
smallest integer such that $\Lambda_{n_\x-1}$ contains the ball of
center $\x$ and radius $\r\, r_\f$.  Then $\forall t\le \Th$

\ifthenelse{\formato=4}{
\be\eqalign{ \widetilde W&\big(\lis x^{(0)}_t, \x, \r \big)\le
\widetilde W\big(\lis x^{(n_\x,0)}_t, \xi,\r \big)\cr &+
|\widetilde W\big(\lis x^{(0)}_t, \x,\r \big)- \widetilde
W\big(\lis x^{(n_\x,0)}_t, \x,\r \big)| \cr} \Eq{e9.29} \ee}
{
\be \widetilde W\big(\lis x^{(0)}_t, \x, \r \big)\le
\widetilde W\big(\lis x^{(n_\x,0)}_t, \xi,\r \big)+
|\widetilde W\big(\lis x^{(0)}_t, \x,\r \big)- \widetilde
W\big(\lis x^{(n_\x,0)}_t, \x,\r \big)| \Eq{e9.29} \ee}
The particles which at any time $t\le \Th$ contribute to Eq.\equ{e9.29}
are in $\L_{n_\x}-1$. Via theorem 6, maximal speed,
distance to the walls and number of interacting particles are bounded
by $C' 2^{n_\x/2}$; hence:

\be|\widetilde W\big(\lis x^{(0)}_t, \x,\r \big)-
\widetilde W \big(\lis x^{(n_\x,0)}_t, \xi,\r \big)|\le C' e^{- c
2^{n_\x/4}}\Eq{e9.30} \ee

Consider first the case of $\r$ large, say $\r> \r_{n_\x}$.
Then $\widetilde W\big(\lis x^{(0)}_t; \x,\r \big)$ can be estimated by
remarking that the argument leading to Eq.\equ{e5.5} remains unchanged
if $R(t)=\r+\int_0^t V_{n_\x}(\t)d\t/\Th$ and $R(t,s)=R(t)+\int_s^t
V_{n_\x}(\t)d\t$ are used instead of the corresponding
$R_{n_\x}(t),R_{n_\x}(t,s)$ (as long as $\r\ge0$).
Then

\be \widetilde W(\lis x^{(0)}_t; \x,\r+ \r_{n_\x}\big)\le C \widetilde
W(x, \r+ 2 \r_{n_\x} \big) \Eq{e9.31} \ee
as in Eq.\equ{e5.5} (see also Eq.\equ{e9.5}).

Suppose $\r_0-\r_{n_\x}> g_{\z}(\x/r_\f)$, hence $\r_0> C n_\x^{\z}$,
then $\widetilde W(\lis x^{(0)}_t;$$ \x,\r_0)\le C' \widetilde
W(x,\r_0 +\r_{n_\x})\le C'' (\r_0+\r_{n_\x})^d\le C \r_0^d$ and
$\NN'\le C \r_0^{d-1} n_\x^{1/2}\le C\r_0^{d-1+1/(2\z)}$.  Hence
$\widetilde W\big(\lis x^{(0)}_t, \x, \r \big)\le
C\,(\r_0^{d}+\r_0^{d-1+1/(2\z)})\le C \r_0^d$ if $\z=1/d$ for $d=2$ and
$\z=1/2$ if $d=3$; likewise one finds $\z=1$ if $d=1$.

The values of $(n_\x-1)^\z\le \r_0\le C n_\x^\z$ are still to be
examined.  In this case, however, the bound $\widetilde W(x^{(0)}_t;
\x,\r+ \r_{n_\x}\big)\le C \widetilde W(x, \r+ 2 \r_{n_\x} \big)$
involves quantities $\r, \r_{n_\x}$ with ratios bounded above and
below by a constant, hence $\widetilde W(\lis x^{(0)}_t; \x,\r) $ is
bounded by $\widetilde W(x; \x,\r+ Cn_\x^{\z})\le C' \r^d $.

Conclusion: there is $C>0$, depending only on $\EE$ and for all
$\r> g_\z(\x/r_\f), \,t\le\Th$ it is $W(x^{(0)}_t; \x,\r)\le C\, \r^d
$ if $\z=1/d$ for $d=1,2$ and $\z=1/2$ if $d=3$.

\*\*
%%%%%%%%%%%%%%%%%%%%%%%%%%%%%%%%%%%%%%%%%%
%%%%%%%%%%%%%%%%%%%%%%%%%%%%%%%%%%%%%%%%%%
\def\SEC{\small Appendix: \AppG}\inizA
%\centerline{\small\bf\piuap\ \  Appendix: \AppG
\section{Free thermostats}\piuap\label{G}
%%%%%%%%%%%%%%%%%%%%%%%%%%%%%%%%%%%%%%%%%%
%%%%%%%%%%%%%%%%%%%%%%%%%%%%%%%%%%%%%%%%%%
\*\*

The need for interaction between particles in order to have a
physically sound thermostat model has been stressed in
\cite{Ru999,Ru001} and provides a measure of the importance of the
problems met above.

The discussion in Sec.\ref{sec4} is heuristic unless the
local dynamics assumption is proved. However if the model is modified
by keeping only the interaction $\f$ between the test particles and
between test particles and thermostat particles, but suppressing
interaction between particles in the same $\O_j,\,j>0$, and,
furthermore, replacing the wall potentials by an elastic collision
rule ({\it i.e.}  supposing elastic collisions with the walls and
$U_j(\V X_j)\equiv 0,\, j>0$) the analysis can be further pursued and
completed. This will be referred as the ``{\it free thermostats}''
model.

In the frictionless case this is the classical
version of the frictionless thermostat models that could be completely
treated in quantum mechanics, \cite{FV963}.

Let $\L_n$ be the ball ${\cal B}(O,2^n)$ of radius $2^n r_\f$ and
$n\ge n_0$ (see the paragraph following Eq.\equ{e2.1}). If $\lis N$
bounds the number of particles in the ball $\BB(O,2^{n_0})$ up to an
arbitrarily prefixed time $\Th$, the first inequality Eq.\equ{e4.2}
and the supposed isoenergetic evolution (which in this case is {\it
also} isokinetic) imply

\be |\a_j|\le N_0 F \sqrt {\frac {\lis N}{2K_j/m}}\le
\frac{N_0F}{\sqrt{d\,k_B T_j/m}}\defi\ell.\Eq{e9.32}\ee
It follows that, for $\z\ge1/d$, the speed of the particles initially in
the shell $\L_{n}/\L_{n-1}$ with radii $ 2^nr_\f,2^{n+1}r_\f$ will
remain within the initial speed up to, at most, a factor
$\l^{\pm1}=e^{\pm \ell \Th}$.  The initial speed of the latter
particles is bounded by, see Eq.\equ{e3.11},

\kern-3mm
\be V_n= v_1 \sqrt{\EE_\z(x)}\,n^{\frac12 \z d} \Eq{e9.33}\ee
Hence, if $n(\Th)$ is the smallest value of $n$ for which the
inequality $2^nr_\f-V_n\,\l\,\Th<D_0+r_\f$ {\it does not hold}, no
particle at distance $> 2^{n(\Th)+1}r_\f$ can interact with the test
system.

This means that $\lis N\le \EE_\z(x)\, 2^{\,(n(\Th)+1)\,d}$ and the
dynamics $x^{(n,a)}(t)$ becomes a finitely many particles dynamics
involving $\le N_0+\lis N$ particles at most.

{}From the equations of motion for the $N_0+\lis N$ particles we see
that their speed will never exceed

\be V_\Th=(\lis V +F\,N_0\,\lis N\,\Th) \,\l\Eq{e9.34}\ee
if $\lis V$ is the maximum of their initial speeds. In turn this means
that for $n$ large enough a better bound holds:

\be|\a_j \dot q_i|\le \frac{N_0 \lis N V_\Th^2 F}{\o_{j,n} 2^{dn}r_\f^d\d\, k_B
  T/m}\tende{n\to\infty}0\Eq{e9.35}\ee
with $T=\min_{j>0}T_j$ and $\d=\min_{j>0}\d_j$ and $\o_{j,n}
2^{dn}r_\f^d$ bounds below (for suitable $\o_{j,n})$ the volume of
$\O_j\cap\L_n$.

Hence, for $a=0,1$, it is $\lim_{n\to\infty}
x^{(n,a)}(t)=x^{(0)}(t)$, and {\it also} the dynamics is local in the
above sense. This completes the analysis of free
thermostats and proves:

\*
\0{\bf Theorem 12:} {\it Free isoenergetic and frictionless thermostats
  are equivalent in the thermodynanic limit}
\*

Notice that essential use has been made of the property
that, in absence of interaction among pairs of thermostat particles
and in presence of perfectly elastic walls, isokinetic and
isoenergetic dynamics coincide: so the denominators in Eq.\equ{e9.32}
are constant.

It would be possible to consider non rigid walls, modeled by a soft
potential $\ps$ diverging near them. We do not do so because it will
be a trivial consequence of the analysis in this paper. The example of
this section has been chosen because it pedagogically illustrates well
the simplest among the ideas of the coming analysis.

%\*
%\0{\it Proof:} Replace $S^{(n,1)}_tx$ by $S^{(n,0)}_tx$
%everywhere in Sec.\ref{sec7}.
\*\*
%%%%%%%%%%%%%%%%%%%%%%%%%%%%%%%%%%%%%%%%%%%%%%%%%%%%%%%%%%%%%%%
\def\SEC{Appendix: \AppH}\label{H}\inizA\piuap
%\centerline{\small\bf\piuap\ \ Appendix: \AppH}
\section{Quasi invariance}
\*\*
%%%%%%%%%%%%%%%%%%%%%%%%%%%%%%%%%%%%%%%%%%%%%%%%%%%%%%%%%%%%%%%

A probability distribution $\m$ on a piecewise regular manifold $M$ is
{\it quasi invariant} for a flow $x\to S_t x$ generated by a
differential equation $\dot x= v_x$ if $e^{-\l(t)}\le
\m(S_{-t} dx)/\m(dx)\le e^{\l(t)}$ and $\l(t)<\infty$.

Suppose given $\Th>0$, a piecewise smooth surface $\Si\subset M$ with
unit normal vector $n_x$ and a ``stopping time'' $x\to\th(x)\le \Th$
defined on $\Si$ consider all points $x\in\Si$ which are reached {\it
for the first time} in positive time $t\le \th(x)$ from data
$y\not\in\Si$. Call $E$, the set of such points, {\it i.e} the {\it
tube with base $\Si$ and ceiling $\th(x)$}.

The probability distribution $\m$ is {\it quasi invariant} with
respect to $\Si$ and to the stopping time $x\to\th(x)$ if it is
absolutely continuous with respect to the volume measure, its density
$r(x)$ is continuous and $e^{-\l}\le \m(S_{-t}dx)/\m(dx)\le e^{\l}$
for some $\l>0$ and for all $0\le t\le \th(x)$: this is referred to by
saying that $\m$ is quasi invariant with respect to the stopping time
$\th(x)$ on $\Si$: symbolically $\m$ is $(\Si,\th(x))$--$\l$-quasi
invariant.

Let $ds_x$ be the surface element on $\Si$ and $\n(d s_x)$ its area
computed by measuring the
volume with $\m$.  Then the following {\it Sinai's lemma},
\cite{Si974,MPP975,FD977}, holds: \*

\noindent{\bf Lemma 3:} {\it If $\m$ is $(\Si,\th(x))$--$\l$-quasi
  invariant the integral of any non negative function $f$ over the
  tube with base $\Si$ and ceiling $\th(x)$ can be bounded by

$$\eqalignno{
\int_E &f(y)\m(dy) \le e^\l \int_\Si \int_0^{\th(x)}  f(S_{-\t}x)
\,|v_x\cdot n_x|\, \n(ds_x)d\t,\cr
&
\ge e^{-\l} \int_\Si \int_0^{\th(x)} f(S_{-\t}x)
\,|v_x\cdot n_x|\, \n(ds_x)d\t&{\rm\eq{e9.36}}\cr}$$
}
\*

The lemma can been used to reduce dynamical estimates to equilibrium
estimates.
\*

\noindent{Proof:} Let the trajectory of a point $y$ which reaches
$\Si$ within the stopping time at $x\in\Si$ be parameterized by the
time $\t$. Then the set of points into which the parallelepiped $\D$
with base $ds_x$ and height $d\t$ evolves becomes a parallelepiped
$S_{t}\D$ with base $S_t ds_x$ and the {\it same height}
$d\t$. Therefore the measure of $\m(S_t\D)$ is $e^{-\l} \le
\frac{\m(S_t\D)}{\m(\D)} \le e^\l$ hence the integral of any positive
function $f(y)$ over the set $E$ can be bounded above and below by the
integral of $\int_{\Si}\int _0^{\th(x)} f( S_{-t} x) \r(x) \n(ds_x) d\t$
if $\r(x)\n(ds_x) d\t$ is the measure of $\D$: the latter is
$|v_x\cdot n_x| \n(ds_x)d\t$. Notice that $\r(x)=|r(x)\,v_x\cdot n_x|$
if $\m(dx)=r(x) dx$.  \*\*

%%%%%%%%%%%%%%%%%%%%%%%%%%%%%%%%%%%%%%%%%%%%%%%%%%%%%%%%%%%%%%%
\def\SEC{\small Appendix: \AppI}\label{I}\inizA\piuap
%\centerline{\small\bf\piuap\ \ Appendix: \AppI}
\section{Regularized thermostatted dynamics}
\*\*
%%%%%%%%%%%%%%%%%%%%%%%%%%%%%%%%%%%%%%%%%%%%%%%%%%%%%%%%%%%%%%%

Consider $N$ particles in $\cup_{j}\O_j\cap\L$ with a configuration of
immobile particles outside $\L$. The analysis in \cite{MPPP976} can be
followed and the solution of the equations of motion can be defined on
the set $\G^+$ consisting of the configurations $x$ in which $1$ of
the particles is at $\x$ on the boundary $\partial\L$, where elastic
collisions take place, with normal speed $\dot q\cdot n(\x)>0$. The
time evolution makes sense until the time $\t_+(x)$ of next collision;
it can then be continued after the elastic collision because, apart
from a set of zero volume, the normal speed of the collision can be
assumed $\ne0$, until the time $t_\L(x)>0$, {\it if any}, in which the
total kinetic energy in one of the containers $\O_j$, $j\ge0$,
vanishes.

Consider the subset $\G(\Th)\subset \G^+$ of the points with
$\t(x)<\Th$. If $\D\subset \G(\Th)$ and if $\n$ is the measure
$\n_{0,\Si'}$, see Eq.\equ{e7.9} and appendix $D$,
is a small set it will be
$\n(T\D)\ge \l^{-1} \n(\D)$ where $\l(x)$ is an upper bound on the
entropy production in any interval within $[0,\t(x)]$.

There cannot be infinitely many collisions, except on a set of $0$
$\n$--measure, as long as $\l(T^n x)<L<\infty$. If $T$ denotes the
return map the only way an
accumulation of collision times with $\sum_{k=0}^\infty \t(T^k
x)=\th<\Th$ could occur is if also $K(S^{(n,1)}_t x)$ $\tende{t\to\th}0$
and $\sup_{t',t''\le \Th} $ $\int_{t'}^{t''} \s(S^{(n,1)}_t x) dt=+\infty$:
hence the limit $\lim_{t\to\th}$ $K(S^{(n,1)}_tx)=0$. The reason is that
if $\l$ remains finite then Poincar\'e's recurrence argument will
apply implying, as in \cite{MPPP976}, that $\sum_{k=0}^\infty \t(T^k
x)=+\infty $ outside a set of $\n$--measure $0$.

This means that the  thermostatted time evolution is well defined until
the first time $t_\L(x)$ when $K(S^{(n,1)}_{t_\L}x)=0$. Since the
thermostat force $\frac{Q_j}{K_j}\dot q_i$ is bounded even if
$K_j\to0$ the time evolution will be defined in the closed interval
$[0,t_{\L}(x)]$: {\it the evolution proceeds, well defined, until the first
time $t_\L(x)$ (included, if any) when some of the $K_{\L,j}$
vanishes}, see remark (6) p.\pageref{t}.

\*\*
%\ifthenelse{\formato=4}{\vfill\eject}{}
%%%%%%%%%%%%%%%%%%%%%%%%%%%%%%%%%%%%%%%%%%%%%%%%%%%%%%%%%%%%%%%
\def\SEC{Appendix: \AppJ}\label{J}\inizA\piuap
%\centerline{\small\bf\piuap\ \ Appendix: \AppJ}
\section{Entropy bounds: check of Eq.\equ{e7.7}}
\*\*
%%%%%%%%%%%%%%%%%%%%%%%%%%%%%%%%%%%%%%%%%%%%%%%%%%%%%%%%%%%%%%%

Writing $k_\xi$ for the smallest integer $\ge (\log n)^\g
g_\g(\xi/r_\f)$ (here $g_\g$ is chosen instead of the natural
$g_{1/2}$ in order to simplify the formulae: recall that by
defiinition $\g>1/2$), then $\mu_{0,\Si'}$ almost surely, $\Si'$
splits into an union over $\xi \in \Lambda_n\cap r_\f\ZZZ^d$ of the
union of $\mathcal S^1_\xi\cup \mathcal S^2_\x\cup{\cal S}^3$, where

\be\eqalign{ \mathcal
S^1_\xi=&\{y\in \Si':|y\cap C_\xi|=k_\xi, |y\cap
\partial C_\xi|=1\}\cr
\mathcal S^2_\xi=&\{y\in \Si':y\cap C_\xi\ni
(q,\dot q), \e(q,\dot q)=\wt \e_\x\}\cr
{\cal S}^3=& \{y\in \Si':
K_{j,n}(y = \frac12\k 2^{nd}\}\cr}
\Eq{e9.37} \ee
if $\wt\e_\xi\defi \big((\log n)^\g g_{\g}(\xi/r_\f)\big)$.

Consider first the case of ${\cal S}^3$.
Let $D\subset {\cal S}^3$ be the set of the $x$ which satisfy
$K_{j,\L_n}(x)=\frac{\f_0}2 \k 2^{dn}$ for a given $j>0$ while
$K_{j',\L_n}(x)>\frac12 \k 2^{dn}$ for $j'>0,\,j'\ne j$.

Recall the DLR-equations, \cite{LR969}, and consider the classical
superstability estimate on the existence of $c>0$ such that $p_n=
e^{-c 2^{dn}}$ bounds the probability of finding more than $\r 2^{d
n}$ particles in $\L_n\cap\O_j$ if $\r$ is large enough ({\it e.g.}  $\r>
\max_j\d_j$). Then the probability $\m_0(D)$ can be bounded by
$p_n$ (summable in $n$) plus, see Appendix H,

\ifthenelse{\formato=4}{
$$\eqalignno{ & e^{CMV}\int \lis\m_0(dq')\sum_{\ell=1}^{\r 2^{n d}}
\Th\int \frac{e^{-\b_j( U_{\L_n,j}(q,q')-\l_j\,\ell)}}{Z_{\L_n,j}(q')}
\frac{dq} {\ell!}\cr &\cdot\, e^{-{\b_j P^2}}\, \widehat
P\,P^{\ell\,d-1 } \,\frac{\o(\ell\,d)}{(\p /\b)^{\ell
d/2}}&\eq{e9.38}\cr}$$}
{\be e^{CMV}\int \lis\m_0(dq')\sum_{\ell=1}^{\r 2^{n d}}
\Th\int \frac{e^{-\b_j( U_{\L_n,j}(q,q')-\l_j\,\ell)}}{Z_{\L_n,j}(q')}
\frac{dq} {\ell!}\cdot\, e^{-{\b_j P^2}}\, \widehat
P\,P^{\ell\,d-1 } \,\frac{\o(\ell\,d)}{(\p/ \b)^{\ell
2d}}\Eq{e9.38}\ee}
where $\lis\m_0$ is the distribution of the positions in $\m_0$,
$q=(q_1,\ldots,q_l)\in (\O_j\cap\L_n)^l$, $P^2=\frac12 \k
2^{dn}$, $U_{\L_n,j}(q,q')$ is the sum of $\f(q-q')$ over pairs of
points $q_i,q'_\ell\in \O_j\cap\L_n$ and pairs
$q_i\in\L_n\cap \O_j, q'_\ell\not\in\L_n\cap \O_j $, and
\*

\noindent(1) $Z_{\L_n,j}(q')$ is the partition function for the region
$\L_n\cap\O_j$ (defined as in Eq.\equ{e3.1} with the integral over the
$q$'s extended to $\L_n\cap\O_j$ and with the energies
$U_{\L_n}(q,q')$);
\*\noindent(2) The volume element $P^{ld-1}dP$ has been changed
to $P^{ld-1}\dot P d\t\defi P^{ld-2} \widehat P d\t$ where $\widehat P$ is
a {\it short hand} for

\be\eqalign{
&\sum_{q,q';\, q\in\L_n\cap\O_j} |\partial_q
\f(q-q')|+\sum_{q\in \L_n\cap\O_j} |\partial_q\ps(q)|
\cr
}\Eq{e9.39}\ee
so that $P\widehat P$ is a bound on the time derivative $2P\dot P$ of
the total kinetic energy $P^2$ contained in $\L_n$ evaluated on the
points of $D$: hence $2P\widehat P$ is bounded by $\le C P^2\widehat
P$).
\*\noindent(3) $\o(\ell\,d)$ is the surface of the unit ball in
$\RRR^{\ell\,d}$.
\*\noindent(4) The $e^{CMV}$ takes into account
the entropy estimate {\it i.e.} the bound, $CMV$, of the
non-invariance of $\m_0$.   \*

The integral can be (trivially) imagined averaged over an auxiliary
parameter $\e\in[0,\lis\e]$ with $\lis\e>0$ arbitrary (but to be
suitably chosen shortly) on which it does not depend at first. Then if
$P$ is replaced by $(1-\e)P$ in the exponential while $P^{\ell\,d-1}$ is
replaced by $\frac{((1-\e)P)^{\ell\,d-1}}{(1-\e)^{d\r 2^{nd}-1}}$ the
average over $\e$ becomes an upper bound. Changing $\e$ to $P\e$ ({\it
i.e.} hence $d\e$ to $({\frac{2}{\k 2^{dn}}})^{\frac12}dP\e$) the bound
becomes the $\m_0$-average $\frac{2 e^{\lis\e \r 2^{nd}}}{\lis\e \k
2^{dn}}\langle {\widehat P\, \ch_{\k,\lis\e}} \rangle_{\m_0}$ bounded
by

\be
\frac{2^{\frac12}e^{\lis\e C 2^{nd}}}{\lis\e \k^{1/2} 2^{dn/2}}
\langle \widehat P^2\rangle^{1/2}_{\m_0}\,
\cdot\,\langle \ch_{\k,\lis\e} \rangle^{1/2}_{\m_0}\le C\, e^{-c\,
2^{{nd}/2}}\Eq{e9.40} \ee
where $\ch_{\k,\e}$ is the characteristic function of the set $\big\{
(1-\lis\e)^2\frac\k2 2^{dn}<K_j<\frac\k2 2^{dn}\big\}$. The inequality is
obtained by a bound on the first average, via a superstability
estimate, proportional to $2^{2dn}$ and by the remark that the second
average is over a range in which $K$ shows a large deviation from its
average (by a factor $2$) hence it is bounded above by $e^{-b 2^{nd}}$
with $b$ depending on $\k$ but independent on $\lis\e$ for $n$
large. Therefore fixing $\lis\e$ small enough (as a function of $\k$)
the bound holds with suitable $C,c>0$ and is summable in $n$ (and of
course on $j>0$).

Similarly, the surface areas $\n_{0,\Si'}({\cal S}^1_\x)$ and
$\n_{0,\Si'}({\cal
S}^2_\x)$ on ${\cal S}^1,{\cal S}^2$ induced by $\m_0$ are bounded by

\be \n_{0,\Si'}({\cal S}^i_\x)\le C e^{C MV}\sqrt{n} e^{-c
[(\log n)^\l g_\l(\x/r_\f)]^2 }, \Eq{e9.41}\ee
(for suitable $C,c$, functions of $E$).  Summing (as $\g>1/2$) over
$\x\in\O_j\cap\L_n$ as discussed below and adding Eq.\equ{e9.38} the
Eq.\equ{e7.7} will follow.

Consider  the case of ${\cal S}^1_\x$. By Eq.\equ{e9.37} if $y\in
\mathcal S^1_\x$ then $|y\cap C_\xi|=k_\xi$ and there is $(q,\dot
q)\in y$ with $q\in \partial C_\xi$.

Remark that $y$ is the configuration reached starting from an initial
data $x\in \XX_E$ within a time $T_{n}(x)<\Th$: hence
Eq.\equ{e5.8} applies. By Eq.\equ{e5.8}

\ifthenelse{\formato=4}{
\be\eqalign{
&\int_{\mathcal S^1_\xi}  \mu_{0,\Si'}(dy)\int_{0}^{\theta(y)}dt\,w(y)\,
\cr
&\le \Theta v_1 Cn^\h\int \mu(dx) \frac
{J_1}{Z_{C_\xi}(x)}\cr}\Eq{e9.42} \ee}
{\be
\int_{\mathcal S^1_\xi}  \mu_{0,\Si'}(dy)\int_{0}^{\theta(y)}dt\,w(y)\,
\le \Theta v_1 Cn^\h\int \mu(dx) \frac
{J_1}{Z_{C_\xi}(x)}\Eq{e9.42} \ee}
where  $\mu(dx)$ is the  $\mu_0$-distribution
of  configurations $x$ {\it outside} $C_\xi$ and

\be J_1= \int_{\partial C_\xi} dq_1 \int_{C_\xi^{k_\xi-1}}
\frac{dq_2\dots dq_{k_\xi}}{(k_\xi-1)!}  \int_{\rrr^{dk_\xi}} d\dot q
e^{-\beta_j H(q,\dot q|x)}\Eq{e9.43}\ee
The estimate of the {\it r.h.s.} of Eq.\equ{e9.1}, as remarked, is an
``equilibrium estimate''.  By superstability, \cite{Ru970}, and since
$\f\ge0$, the configurational energy $U(q|x)$ $\ge bk_\xi^2
-b'k_\xi$, so that $J_1$ is bounded by:

 \be B\,e^{-\beta_j(bk_\xi^2 -b'k_\xi)} \frac{|C_\xi|^{k_\xi-1}|
\partial C_\xi|}{(k_\xi-1)!} (\frac{2\pi } {\beta_jm})^{\frac{d}{2}
{k_\xi}}\Eq{e9.44}\ee
while $\int \mu(dx) \frac 1{Z_{C_\xi}(x)}\le 1$ because
$Z_{C_\xi}(x)\ge 1$: and the bound can be summed over $k_\x$.  Thus
the contribution from ${\cal S}^1_\x$ to Eq.\equ{e7.7} is bounded by

\be C' e^{C M V }n^\h\, e^{-b [(\log n)^\g g_\g( \x
     /r_\f)]^{2}}\Eq{e9.45}\ee
with $C,b$ suitable positive constants.  Since $\g>1/2$, this is
summable over $\x$ and yields the part of the Eq.\equ{e7.7} coming
from the integration over ${\mathcal S^1}$.

Let, next, $y\in {\cal S}^2_\xi$ and let $(q,\dot q)$ as in
\equ{e9.37}. The function $w$ is, if $E(q,\dot q)\equiv
\e(q,\dot q)^2$,

\be w = \frac{|d E(q,\dot q)/dt|}{|{\rm
 grad}E(q,\dot q)|} \Eq{e9.46}\ee
and $|d E(q,\dot q)/dt| \le C\,|\dot q| n^\h$ because $d E/dt$ is the
work on the particle $(q,\dot q)$ done by the pair interactions
(excluding the wall forces). It is bounded proportionally to the
number of particles which can interact with $(q,\dot q)$, which, by
theorem 4, is bounded proportionally to $n^{1/2}$ (as the total
configuration is in $\Si'$). On the other hand, $|{\rm grad}E(q,\dot
q)| = \sqrt{|\partial \psi(q)|^2 + m^2|\dot q|^2}\ge {m}|\dot q|$
hence $w\le C n^{1/2}$ again by Eq.\equ{e5.8} and the remark
preceding Eq.\equ{e9.42}.

Then, analogously to \equ{e9.42}, the integral under consideration is
bounded by $C e^{C' MV }\sqrt{n}$ ($C,C'$ are suitable constants
functions of $E$) times an equilibrium integral $\int \mu(dx) \frac
{J_2}{Z_{C_\xi}(x)}$
with $J_2$ defined by:

\ifthenelse{\formato=4}{
$$\eqalignno{ \sum_k & \int_{C_\xi^{k-1}\times \rrr^{k-1}}
  \frac{dq_2\dots dq_k d\dot q_2\ldots d\dot q_{k}}{(k-1)!}e^{-\b_j
  H(q,\dot q|x)+\b_j\l_jk}\cr & \cdot e^{-\beta_j \tilde E_\xi}
  \,{\rm area}(\{E(q,\dot q)=\tilde E_\x\})&\eq{e9.47} \cr} $$}
{\be\sum_k \int_{C_\xi^{k-1}\times \rrr^{k-1}}
  \frac{dq_2\dots dq_k d\dot q_2\ldots d\dot q_{k}}{(k-1)!}e^{-\b_j
  H(q,\dot q|x)+\b_j\l_jk}\cdot e^{-\beta_j \tilde E_\xi}
  \,{\rm area}(\{E(q,\dot q)=\tilde E_\x\})\Eq{e9.47}\ee}
\*

\noindent{}where the ${\rm area} (\{E(q,\dot q)=\tilde E_\x\})$ is the
area of the surface $\{(q,\dot q):E(q,\dot q)=\tilde E_\xi\}$ in
$\RRR^{2d}$ (the $\wt E_\x$ is defined in \equ{e9.37}).  Then $J_2$ is
bounded by

\be\textstyle\sum_k \frac{B}{(k-1)!} \Big(e^{\b_j\l_j}|C_\xi| \big(\frac{2\pi }
       {\beta_jm}\big)^{\frac{d}2}\Big)^{(k-1)} |C_\xi| (\tilde
       E_\xi)^{(d-1)/2} e^{-\beta_j \tilde E_\xi} \Eq{e9.48}\ee
so that, suitably redefining $C,C'$ (functions of $E$), the
contribution from ${\cal S}^2_\x$ is bounded by

\be C' e^{C M V }\sqrt{n} e^{-\frac{1}{2}\beta_j [(\log n)^\g g_\g(
\x /r_\f)]^{2}} \Eq{e9.49} \ee
and Eq.\equ{e7.7} follows from Eq.\equ{e9.38}, from \equ{e9.45} and
from Eq.\equ{e9.49}.

\*\*
%%%%%%%%%%%%%%%
\def\SEC{\small Appendix: \AppK}\label{K}\inizA\piuap
%\centerline{\small\bf\piuap\ \ \AppK}
\section{Details the derivation of Eq.\equ{e7.5},\equ{e7.6}}
\*\*
%%%%%%%%%%%%%

To prove item (1) and Eq.\equ{e7.6}, thus completing the proof of
theorem 7, we shall compare the evolutions $x^{(n,1)}(t)$ with $\lis
x^{(n,0)}(t)$, at same initial datum $x\in \XX_E$ and $t\le T_{n}(x)$,
the latter being the stopping time defined in Eq.\equ{e7.4}. We start
by proving that there is $C>0$ so that for all $n$ large enough the
following holds.
\*

\noindent{\bf Lemma 4:} {\it For $t\le T_n(x)$, see Eq.\equ{e7.4},
  and $k\ge (\log n)^\g$, then
\ifthenelse{\formato=4}{
\be\eqalign{
|\dot q^{(n,1)}_i(t)|&\le C\,v_1\, \big(k\,\log n)^\g,
\cr
|q^{(n,1)}_i(t)|&\le r_\f\,(2^k+C \, \big(k\,\log n)^\g).
\cr}
\Eq{e9.50}\ee}
{
\be
|\dot q^{(n,1)}_i(t)|\le C\,v_1\, \big(k\,\log n)^\g,
\qquad
|q^{(n,1)}_i(t)|\le r_\f\,(2^k+C \, \big(k\,\log n)^\g).
\Eq{e9.50}\ee}
}
for $q_i(0)\in\L_k$.
\*

The necessity of this lemma is due to the fact that we cannot control
the positions and speed at time $t$ in terms of the norms $\|x\|_n$ at
time $0$: since the particles move we must follow them (a
``Lagrangian'' viewpoint).

A corollary of the above will be:
 \*

\0{\bf Lemma 5:} {\it Let $\NN$ and $\r$ be the maximal
  number of particles which at any given time $\le T_n(x)$ interact with a
  particle $q_i$ initially in $\L_{k+1}$ and, respectively, the minimal
  distance of a particle from the walls. Then

\be
\NN \le C\,(k \log n)^{d\g},\;\; \r \ge\, c\, (k \log n)^{-2(d\g+1) /\a}
  \Eq{e9.51}\ee
for all integers $k\in((\log n)^\g, 2(\log n)^\g)$.}
\*

The proof of the lemmas is in Appendix L. The interval allowed to $k$
could be much larger, as it appears from the proof.
\*

We have now all the ingredients to bound $\d_i(t,n) \defi$ $
|q_i^{(n,1)}(t)- \lis q_i^{(n,0)}(t)|$.  Let $f_i$ be the acceleration
of the particle $i$ due to the other particles and to the walls.  In
the equations of motion for the evolution, Eq.\equ{e3.7}, for
$q^{(n,1)}_i(t)$ the elastic collisions with $\partial\L_n$ (implicit
in Eq.\equ{e3.7}) can be disregarded if the particle initially in
$q_i$ does not experience a collision with $\partial\L_n$.  For $n$
large such a collision cannot take place if $q_i(0)\in\L_{k+1}$,
$k<2(\log n)^\g$ and $t\le T_{n}(x)$, by Eq.\equ{e9.50}.

Then, for $2(\log n)^\g>k>(\log n)^\g$, by Eq.\equ{e9.51} and if $q_i\in
\L_{k+1}$, it follows that $|f_i|\le C\, (k\log n)^{\h}$, with $\h$
a suitable constant ($\h\defi d\,\g\,+2(d\,\g+1)(1+\frac 1\alpha)$),
so that subtracting the Eq.\equ{e3.7} and Eq.\equ{e6.1} for the two
evolutions, it follows that for any $q_i\in \L_{k+1}$ (possibly close
to the origin hence very far from the boundary of $\L_k$, if $n$ is
large)

\ifthenelse{\formato=4}{
\be\eqalign{ &\d_i(t,n)\le  C \, (k\log n)^{\h} 2^{-nd}
\cr& + \Th
\int_0^t |f_i(q^{(n,1)}(\t))-f_i(\lis q^{(n,0)}(\t))|\,d\t,\cr
}\Eq{e9.52}\ee
}
{\be\d_i(t,n)\le  C \, (k\log n)^{\h} 2^{-nd}
+ \Th
\int_0^t |f_i(q^{(n,1)}(\t))-f_i(\lis q^{(n,0)}(\t))|\,d\t,
\Eq{e9.52}\ee}
because, recalling the definition of $\a_j$ and the third condition on
the stopping time in Eq.\equ{e7.2}, the denominator of $|\a_j|$ is bounded
proportionally to $2^{nd}$.

Remarking, by lemma 4 for $q^{(n,1)}_i(t)$, that

 \be \max_{t\le T_{n}(x)} |q_i^{(n,1)}(t)-q_i| \le C\, r_\f (k
\log n)^{\g},\; \Eq{e9.53}\ee
and, by theorem 5 for $\lis q_i^{(n,0)}(t)$), also $\lis
q_i^{(n,0)}(t)$ satisfies an identical relation, let $\ell>0$ be an
integer, $k_\ell$ such that
\be 2^{k_\ell}=2^k +\ell \,C\, (k \log n)^{\g}. \Eq{e9.54}\ee
and let $u_{k_\ell}(t,n)$ the max of $\d_i(t,n)$ over $|q_i|\le
r_\f 2^{k_\ell}$.  Then by Eq.\equ{e9.52} and Eq.\equ{e9.51},

\be
\frac{u_{k_\ell}(t,n)}{r_\f}\le \,C\, (k\log
  n)^{\h} \,\big
(2^{-nd} +\int_0^t\frac{ u_{k_{\ell+1}}(s,n)ds}{r_\f\Th}\big),
\Eq{e9.55}\ee
for $\ell \le \ell^*=2^k/((k\log n)^\g C)$, the latter being the
largest $\ell$ such that $2^{k_\ell}\le 2^{k+1}$.  By
Eq.\equ{e9.55},\equ{e6.10}, and Eq.\equ{e9.51}

\ifthenelse{\formato=4}{
$$\eqalignno{ & \frac{u_{k}(t,n)}{r_\f} \le e^{C\, (k\,\log n)^{\h}} C
(k\log n)^{\h} 2^{-dn}
&\eq{e9.56}
\cr& + \frac{(C\, (k\,\log
    n)^{\h})^{\ell^*}} {\ell^*!}\, C\,(2^k+k (\log
    n)^{\g}+k^{1/2}),\cr }$$}
{
\be\frac{u_{k}(t,n)}{r_\f} \le e^{C\, (k\,\log n)^{\h}} C
(k\log n)^{\h} 2^{-dn}
+ \frac{(C\, (k\,\log
    n)^{\h})^{\ell^*}} {\ell^*!}\, C\,(2^k+k (\log
    n)^{\g}+k^{1/2}),\Eq{e9.56}\ee}
for $(\log n)^\g<k<2(\log n)^\g$.

Thus $u_{k}(t,n)$ is bounded by the r.h.s.\ of the
first of Eq.\equ{e7.5}; analogous argument shows that also the
velocity differences are bounded as in Eq.\equ{e7.5} which is thus
proved for all $t\le T_{n}(x)$.

Therefore given $q_i(0)$ with $|q_i(0)|/r_\f\le 2^{k_0}$ it is, for
$n> e^{k_0^{1/\g}}$ large enough and $i$ fixed, $|q_i^{(n,1)}(t)-\lis
q^{(n,0)}_i(t)|/r_\f<u_{(\log n)^\g}(t,n)\le C\, 2^{- c \, (\log n)^\g}$,
{\it i.e.} for $n$ large $q^{(n,1)}_i(t)$ is closer than $r_\f$ to
$\lis q^{(n,0)}_i(t)$.

It remains to check Eq.\equ{e7.6}: for $n$ large the number of
particles in $x^{(n,1)}_i(t)$ which are in $\L_*$ is smaller than the
number of particles of $\lis x^{(n,0)}_i(t)$ in $ \L_{**}$ which in
turn is bounded by $M$ up to time $\Th$, by theorem 6.  An analogous
argument for the velocities allows to conclude Eq.\equ{e7.6}.

Applying again Eq.\equ{e7.5} the proof of theorem 7 is complete apart
from the uniqueness statement: the latter can be proved (as usual by
iteration of the equations in integral form) but we skip the details,
see \cite{CMP000,CMS005}. The uniqueness is within the solutions of the
equations of motion in the infinite volume and with data subect to the
condition of being in $\HH_{1/d}$.

\*\*
%%%%%%%%%%%%%%%
\def\SEC{Appendix: \AppL}\label{L}\inizA\piuap
%\centerline{\small\bf\piuap\ \ \AppL}
\section{Proof of Lemmas 4,5}
\*\*
%%%%%%%%%%%%%

With the notations of Appendix K and $\g>\frac12$:
\*
\0{\it Proof (lemma 4):} If $t\le T_{n}(x)$ then
\be \kern-3mm|\dot q^{(n,1)}_i(t)|\le v_1 \big((\log n) \log_+ \frac{
  |q^{(n,1)}_i(t)|+\sqrt2 r_\f}{r_\f} \big)^\g,\Eq{e9.57}\ee
implying: $|q^{(n,1)}_i(t)|\le r(t) r_\f$ if $r(t)\,r_\f$ is an upper
bound to a solution of Eq.\equ{e9.57} with $=$ replacing $\le$ and
initial datum $|q^{(n,1)}_i(0)| \le 2^kr_\f$.  And $r(t)$ can be taken
$r(t)\defi 2^k+ 2v_1 \big((\log n) \log_+ { 2^k}\big)^\g \,
\frac{t}{r_\f}$, for $t\le \Theta$, provided

\be  \kern-3mm\big((\log n) (\log_+{ r(\Theta)+\sqrt2})\big)^\g \le 2
\big((\log n) \log_+{ 2^k}\big)^\g \Eq{e9.58}\ee
which, since $\frac{(k\log n)^\g}{2^k}$ vanishes as $n$ diverges
  (because $k\ge (\log n)^\g$), is verified for all $n$ large enough.
  Thus $|q^{(n,1)}_i(t)|\le \,r_\f\,r(t)$, hence $|\dot
  q^{(n,1)}_i(t)|\le \, r_\f\, C\,\dot r(t)$ for all $t\le
  T_{n}(x)$, and the
  lemma is proved.  \*

\0{\it Proof (lemma 5):} By lemma 1 the following
properties hold for all $n$ large enough and all $t\le
T_{n}(x)$:  \*

\noindent
(i) for all $q_i\in \L_{k+2}$,
 \be \max_{t\le T_{n}(x)} |q_i^{(n,1)}(t)-q_i| \le C\, r_\f (k
\log n)^{\g},\; \Eq{e9.59}\ee
which implies the first of Eq.\equ{e9.51}.
\\
(ii) particles in $\L_{k}$ do not interact with those
$\not\in\L_{k+2}$; \*

By Eq.\equ{e9.59} we see that if $q_i\in\L_{k+1}$ then
$q_i^{(n,1)}(t)\in \L_{k+2}$ so that, by the definition of the set
$\BB$, recalling that $\g>1/2$, the lemma
follows because the work of the pair forces on $q_i$ is bounded
$\NN\le C (k\log n)^{d\g}$ times the speed bound $C (k \log n)^\g$,
by Eq.\equ{e9.50} and the second of \equ{e9.51}. \*\*

\ifthenelse{\formato=4}{\noindent
{\bf Acknowlegements}: This work has
been partially sup\-por\-ted also by Rutgers University.
We are grateful to Dr. S. Simonella for his
very careful comments.
}{}
\*\*

%\newcount\biblio\biblio=0
%  \ifnum\biblio=0\bibliography{0Bib}\fi \ifnum\biblio=1\input TT.bbl
%  \fi%
%\small \bibliographystyle{apsrev}

%\bibliography{0Bib}

\begin{thebibliography}{10}

\bibitem{Ga008d}
G.~Gallavotti.
\newblock On thermostats: {I}sokinetic or {H}amiltonian? finite or infinite?
\newblock {\em Chaos}, 19:013101 (+7), 2008.

\bibitem{FV963}
R.P. Feynman and F.L. Vernon.
\newblock The theory of a general quantum system interacting with a linear
  dissipative system.
\newblock {\em Annals of Physics}, 24:118--173, 1963.

\bibitem{EM990}
D.~J. Evans and G.~P. Morriss.
\newblock {\em Statistical Mechanics of Non{\-}equilibrium Fluids}.
\newblock Academic Press, New-York, 1990.

\bibitem{GG007}
P.~Garrido and G.~Gallavotti.
\newblock Boundary dissipation in a driven hard disk system.
\newblock {\em Journal of Statistical Physics}, 126:1201--1207, 2007.

\bibitem{FD977}
J.~Fritz and R.L. Dobrushin.
\newblock Non-equilibrium dynamics of two-dimensional infinite particle systems
  with a singular interaction.
\newblock {\em Communications in Mathematical Physics}, 57:67--81, 1977.

\bibitem{MPP975}
C.~Marchioro, A.~Pellegrinotti, and E.~Presutti.
\newblock Existence of time evolution for $\nu$ dimensional statistical
  mechanics.
\newblock {\em Communications in Mathematical Physics}, 40:175--185, 1975.

\bibitem{MPPP976}
C.~Marchioro, A.~Pellegrinotti, E.~Presutti, and M.~Pulvirenti.
\newblock On the dynamics of particles in a bounded region: A measure
  theoretical approach.
\newblock {\em Journal of Mathematical Physics}, 17:647--652, 1976.

\bibitem{CMP000}
E.~Caglioti, C.~Marchioro, and M.~Pulvirenti.
\newblock Non-equilibrium dynamics of three-dimensional infinite particle
  systems.
\newblock {\em Communications in Mathematical Physics}, 215:25--43, 2000.

\bibitem{Ru970}
D.~Ruelle.
\newblock Superstable interactions in classical statistical mechanics.
\newblock {\em Communications in Mathematical Physics}, 18:127--159, 1970.

\bibitem{Ga006c}
G.~Gallavotti.
\newblock Entropy, thermostats and chaotic hypothesis.
\newblock {\em Chaos}, 16:043114 (+6), 2006.

\bibitem{Ga000}
G.~Gallavotti.
\newblock {\em Statistical Mechanics. A short treatise}.
\newblock Springer Verlag, Berlin, 2000.

\bibitem{Ga008c}
G.~Gallavotti.
\newblock Thermostats, chaos and {O}nsager reciprocity.
\newblock {\em Journal of Statistical Physics}, 134:1121--1131, 2009.

\bibitem{CMS005}
G.~Cavallaro, C.~Marchioro, and C.~Spitoni.
\newblock Dynamics of infinitely many particles mutually interacting in three
  dimensions via a bounded superstable long-range potential.
\newblock {\em Journal of Statistical Physics}, 120:367--416, 2005.

\bibitem{Si974}
{Ya.}~G. Sinai.
\newblock The construction of the cluster dynamics of dynamical systems in
  statistical mechanics.
\newblock {\em Moscow University Mathematics Bulletin}, 29:124--129, 1974.

\bibitem{ES993}
D.~J. Evans and S.~Sarman.
\newblock Equivalence of thermostatted nonlinear responses.
\newblock {\em Physical Review E}, 48:65--70, 1993.

\bibitem{Ru000}
D.~Ruelle.
\newblock A remark on the equivalence of isokinetic and isoenergetic
  thermostats in the thermodynamic limit.
\newblock {\em Journal of Statistical Physics}, 100:757--763, 2000.

\bibitem{Ru00b}
D.~Ruelle.
\newblock Natural nonequilibrium states in quantum statistical mechanics.
\newblock {\em Journal of Statistical Physics}, 98:55--75, 2000.

\bibitem{ABGM972}
D.~Abraham, E.~Baruch, G.~Gallavotti, and A.~Martin-L{\"o}f.
\newblock Dynamics of a local perturbation in the {$X-Y$} model ({II}).
\newblock {\em Studies in Applied Mathematics}, 51:211--218, 1972.

\bibitem{Le971}
J.~L. Lebowitz.
\newblock {\em Hamiltonian flows and rigorous results in nonequilibrium
  statistical mechanics}, volume Ed. S.A. Rice, K.F.Freed, J.C.Light of {\em
  Proceedings of the VI IUPAP Conference on Statistical Mechanics}.
\newblock University of Chicago Press, Chicago, 1971.

\bibitem{EPR999}
J.~P. Eckmann, C.~A. Pillet, and L.~Rey Bellet.
\newblock Non-equilibrium statistical mechanics of anharmonic chains coupled to
  two heat baths at different temperatures.
\newblock {\em Communications in Mathematical Physics}, 201:657--697, 1999.

\bibitem{BGGZ005}
F.~Bonetto, G.~Gallavotti, A.~Giuliani, and F.~Zamponi.
\newblock Chaotic {H}ypothesis, {F}luctuation {T}heorem and {S}ingularities.
\newblock {\em Journal of Statistical Physics}, 123:39--54, 2006.

\bibitem{Ru002}
D.~Ruelle.
\newblock How should one define entropy production for nonequilibrium quantum
  spin systems?
\newblock {\em Reviews in Mathematical Physics}, 14:701--707, 2002.

\bibitem{GP009}
G.~Gentile and M.~Procesi.
\newblock Periodic solutions for a class of nonlinear partial differential
  equations in higher dimension.
\newblock {\em Communications in Mathematical Physics}, 289:863--906, 2009.

\bibitem{Ru999}
D.~Ruelle.
\newblock Smooth dynamics and new theoretical ideas in non-equilibrium
  statistical mechanics.
\newblock {\em Journal of Statistical Physics}, 95:393--468, 1999.

\bibitem{Ru001}
D.~Ruelle.
\newblock Entropy production in quantum spin systems.
\newblock {\em Communications in Mathematical Physics}, 224:3--16, 2001.

\bibitem{LR969}
O.~Lanford and D.~Ruelle.
\newblock Observables at infinity and states with short range correlations in
  statistical mechanics.
\newblock {\em Communications in Mathematical Physics}, 13:194--215, 1969.

\end{thebibliography}
\small
\ifthenelse{\formato=4}{\bibliographystyle{unsrt}
}{\bibliographystyle{apsrev}}
\def\SEC{\small References}\label{ref}
\vskip3mm
\ifthenelse{\formato=4}{
\noindent{e-mails: \\
\tt giovanni.gallavotti@roma1.infn.it, \\
 presutti@mat.uniroma2.it}

\hfill Roma, 8 Ottobre 2009%\today
}
{e-mails: \\
\tt giovanni.gallavotti@roma1.infn.it,\
 presutti@mat.uniroma2.it,\rm\hfill Roma, \today}
\end{document}